\documentclass[final,12pt]{clear2025} % Include author names

\usepackage{times}

\usepackage[utf8]{inputenc} % allow utf-8 input
\usepackage[T1]{fontenc}    % use 8-bit T1 fonts
\usepackage{hyperref}       % hyperlinks
\usepackage{url}            % simple URL typesetting
\usepackage{booktabs}       % professional-quality tables
\usepackage{amsfonts}       % blackboard math symbols
\usepackage{nicefrac}       % compact symbols for 1/2, etc.
\usepackage{microtype}      % microtypography
\usepackage{xcolor}         % colors

\usepackage{mathtools} % amsmath with fixes and additions
\usepackage{caption}
\usepackage{subcaption}
\usepackage{xcolor}
\usepackage{xargs}

\newcommand{\fref}[1]{Figure~\ref{#1}}
\newcommand{\tref}[1]{Theorem~\ref{#1}}

\newtheorem{pruning}{Pruning}
\newtheorem{update}{Update}
\newtheorem*{divcon*}{Divide-and-conquer algorithm}

\newcommand{\DAG}{{\cal G}}
\newcommand{\N}{{\cal N}}
\newcommand{\ghnode}{h} % generic dormant node
\newcommand{\vn}{v} % generic node
\newcommand{\vnp}{w} % generic other node
\newcommand{\ghvec}{{\boldsymbol{h}}} % generic dormant vector

\newcommand{\s}[2]{s(#1|#2)} % family score
\newcommand{\smax}[2]{s_{\max}(\{#1\}|{#2})}

\newcommand{\osc}[1]{S(#1)} % order score
\renewcommand{\o}{\vn} % order
\newcommand{\bigO}{\mathcal O} % order
\newcommandx{\lo}[2][2=]{
    \ifthenelse{\equal{#2}{}}
        {\langle #1, [\dots] \rangle}
        {\langle #1, [#2] \rangle}
} % left order
\newcommandx{\ro}[2][2=]{
    \ifthenelse{\equal{#2}{}}
        {\langle [\dots], #1 \rangle}
        {\langle[#2], #1\rangle}
} % right order
\newcommand{\fo}[1]{\langle{#1}\rangle} % full order
\newcommandx{\lro}[3][3=]{
    \ifthenelse{\equal{#3}{}}
        {\langle #1,[\dots], #2 \rangle}
        {\langle #1, [#3], #2 \rangle}
} % r% left and right order

\newcommand{\harpoonr}{\overset{\rightharpoonup}}
\newcommand{\harpoonl}{\overset{\leftharpoonup}}

\newcommand{\vr}[1]{\harpoonl{\vn}_{\! #1:1}}
\newcommand{\vl}[1]{\harpoonr{\vn}_{\! 1:#1}}

\newcommand{\vrp}[1]{\harpoonl{\vnp}_{\! #1:1}}

\newcommand{\vstar}{\vn^{*}}
\newcommand{\Sstar}{S^{*}}

\newcommand{\gbound}[1]{g(#1)} % g bound
\newcommand{\fbound}[1]{f(#1)} % f bound

\newcommand{\Pa}{\mathbf{Pa}}
\newcommand{\T}[1]{\mathcal{T}(#1)}
\newcommand{\M}[1]{\mathcal{M}(#1)}

\title[Polynomial exact discovery for sparse Bayesian networks]{Exact discovery is polynomial for certain\\sparse causal Bayesian networks}

\clearauthor{%
  \Name{Felix L. Rios}\and\Name{Giusi Moffa}\\
  \addr{Department of Mathematics and Computer Science, University of Basel, Basel, Switzerland}
  \AND \Name{Jack Kuipers}\Email{jack.kuipers@bsse.ethz.ch}\\
  \addr{D-BSSE, ETH Zurich, Basel, Switzerland}
}

\begin{document}

\maketitle

\begin{abstract}
Causal Bayesian networks are widely used tools for summarising the dependencies between variables and elucidating their putative causal relationships. By restricting the search to trees, for example, learning the optimum from data is polynomial, but this does not guarantee finding the optimal network overall. Without similar restrictions, exact discovery of the optimum is computationally hard in general and no polynomial results are known. The current state-of-the-art approaches are integer linear programming over the underlying space of directed acyclic graphs, dynamic programming and shortest-path searches over the space of topological orders, and constraint programming combining both. For dynamic programming over orders, the computational complexity is known to be exponential base 2 in the number of variables in the network. We demonstrate how to use properties of Bayesian networks to prune the search space and lower the computational cost, while still guaranteeing exact discovery of the provably optimal network. We also include new path-search and divide-and-conquer criteria.  Without \emph{a priori} constraining the search to certain types of networks, the algorithm completes in quadratic time when the optimum is a matching, and in polynomial time when the optimum belongs to any network class with logarithmically-bound largest connected components. In simulation studies we observe the polynomial dependence for sparse networks and that, beyond some critical value, the logarithm of the base grows with the network density. Our approach then out-competes the state-of-the-art at lower densities. These results therefore pave the way for faster exact causal discovery in larger and sparser networks.  
\end{abstract}


\section{Introduction}\label{sec:intro}

\emph{Causal Bayesian networks} have in the last decades become standard models for describing causal relationships among a set of random variables since their directed edges may provide a powerful handle for complex causal queries. In line with increased general interest in causality, the interest in automatic procedures for learning the structure of Bayesian networks encoding the probabilistic dependencies of causal mechanisms, so-called \emph{causal structure learning} or simply \emph{causal discovery}, has also increased. Although finding the best tree is polynomial \citep{chow1968approximating}, this does not prove that the tree is optimal amongst all networks and hence does not achieve exact discovery of the optimal network. Searching within similarly restricted spaces like networks which can be turned into trees or branchings with a fixed number of edge of vertex deletions \citep{gaspers2015finding, gruttemeier2022learning} is also polynomial. However, exact discovery without such restrictions is in general NP-hard \citep{chickering1996learning,chickering2004large} and no polynomial results are known. 
For this reason, most existing algorithms are based on heuristic or approximate methods, and a plethora of different approaches and implementations exist \citep{scutari2019learns,rios2021benchpress}. These are usually either \emph{score-based}, where each network has a goodness of fit score to the data, \emph{constraint-based}, where networks are learned by independence testing, or \emph{hybrid} combining both aspects. With the recent development of software packages \citep{rios2021benchpress} for using and comparing algorithms, we can expect the pace of advances to continue to increase.

In this paper, we wish to examine the NP-hardness in detail and focus on exact score-based causal discovery.
The current state-of-the-art approaches \citep{malone2018empirical} are integer linear programming (ILP) over the underlying space of directed acyclic graphs (DAGs) \citep{bartlett2017integer,cussens2017bayesian,cussens2020gobnilp}, dynamic programming \citep{koivisto2004exact, silander2006simple} and path searches \citep{yuan2013learning} over the space of topological orders \citep{friedman2003being,teyssier2005ordering}. Finally, constraint programming approaches use both spaces \citep{van2015machine,trosser2021improved}, and compared to dynamic programming, they prune networks that can never be optimal. 

Here, we present an algorithm for exact causal discovery with new pruning rules for sequentially eliminating suborders. Also included are alternative bounds on path searches \citep{yuan2013learning}, computable at low complexity, further allowing us to prove optimality in polynomial time for matchings (graphs with at most one neighbour for each node). Combining our pruning rules with a new divide-and-conquer algorithm, we can extend our polynomial exact discovery to classes of sparse graphs. The rest of the paper is structured as follows. Section~\ref{sec:exact_discovery} reviews order-based exact structure learning.
Section~\ref{sec:pruning} describes our new pruning approach which we use to prove optimality for matchings and certain classes of sparse graphs in polynomial time in Section~\ref{sec:poly}. In Section~\ref{sec:sims} we empirically demonstrate the polynomial dependence of our method, and how the hardness depends on the network density beyond some critical point. %Finally,
Section~\ref{sec:discussion} provides a discussion of our results.

\subsection{Preliminaries}
Let $\DAG=(V,E)$ be a DAG with vertex set $V=\{1,\dots,p\}$ and edge set $E$. 
The vertices correspond to the random variables in a Bayesian network and the edges to the direct dependence between pairs of nodes. Following the factorisation of the joint probability distribution of a Bayesian network as $P(V) = \prod_{i=1}^{p}P(i \mid \Pa_i)$, where $\Pa_i$ are the parents of node $i$, the log-score of a network is
\begin{align}
\sigma(\DAG) = \sum_{i = 1}^{p} \sigma(i, \Pa_i \mid D)
\end{align}
where $\sigma$ is a function of the data $D$ depending only on a node and its parents. Examples include penalised likelihoods, like the BIC, or marginalised likelihoods, like the BDe for categorical data \citep{heckerman1995learning} and the BGe for linear-Gaussian data \citep{geiger2002parameter,kuipers2014addendum}.

The task of exact causal discovery is to find a DAG that maximises the score: $\operatorname*{arg\,max}_\DAG \sigma(\DAG)$. Each DAG belongs to at least one topological \emph{order}, $\vn_{1:p}=\fo{\vn_1,\dots,\vn_p}$, of the nodes where the edges are pointing to the left. The score of a node $\vn_j$ in an order depends on a subset of the nodes further right, referred to as \emph{potential parents}. We denote the maximum possible node score by
\begin{align}
\s{\vn_j}{\vn_{j+1:p}} = \max_{\Pa_{\vn_j} \subseteq \left\{\vn_{j+1:p}\right\}} \sigma(\vn_j, \Pa_{\vn_j} \mid D)
\end{align}
As a consequence, more potential parents for a specific node cannot decrease its score. We consider the setting where finding the optimal parent set for each node in the network can be performed in polynomial time, for example by limiting the in-degree, so that the complexity bottleneck is in the network search which is generally NP-hard \citep{chickering1996learning,chickering2004large}. We assign to each order the score of an optimal DAG consistent with that order
\begin{align} \label{eq:orderscore}
\osc{\o_{1:p}} := \sum_{j=1}^p\s{\o_j}{\o_{j+1:p}}.
\end{align}
Finding an order with an optimal score then provides a maximally scoring DAG \citep{koivisto2004exact,teyssier2005ordering}. Note that different orders of the same set of nodes might give the same score, implying that there might be more than one optimal ordering. Since we only need to find one, we distinguish an optimal order (OO) from an ordered optimal order (OOO) where any neighbouring nodes in an OO with the same score on transposition are in numerical order.

Our task is to find the OOO, and hence a DAG, with the maximum score. We focus on finding the highest-scoring network in a restricted search space where each node has a preselected set of parents, up to total size $K$ and one additional parent outside that set. One way to obtain the search space is with a hybrid approach, starting from a constraint-based skeleton and iteratively improving it with search-and-score \citep{kuipers2022efficient}. We focus on this setting since accessing the relevant scores is highly efficient, while the iterative approach of updating the search space, and subsequent sampling of DAGs \citep{kuipers2022efficient}, has been shown to offer good performance in causal discovery benchmarks \citep{rios2021benchpress} as well as enabling Bayesian analyses.  

We distinguish between two types of \emph{suborders} of length $n$, \emph{left} and \emph{right orders}, denoted as $\vl{n} = \lo{\vn_{1:n}}[\dots]$ and $\vr{n} = \ro{\vn_{n:1}}[\dots]$, respectively. For tidiness, we remove brackets in the subscripts so $\vn_{(a+b):(c+d)} =: \vn_{a+b:c+d}$.
The square bracket notation symbolises the set of the nodes outside the order, $V \setminus \{\vn_{j}\}_{j=1}^n$, which we refer to as \emph{dormant}.
Each suborder represents the set of orders matching the specific ordering of the \emph{visible} (\emph{i.e.} not dormant) nodes, thus for an order $\vn^*$ we may write statements like  $\vn^* \in \ro{\vn_{n:1}}[\dots]$.
The dormant nodes have no internal order. However, when needed we may explicitly label them in the notation as $\ro{\vn_{n:1}}[h]$.
Finally, we call the left- and right-most nodes in a suborder the \emph{front} and \emph{back}, respectively.
\begin{definition}
The score of a (sub)order is the sum of the scores for the individual visible nodes.
\end{definition}
For left orders, say, we truncate the sum in Equation~(\ref{eq:orderscore}) to $j\leq n$. We let $\smax{A}{B}$ denote the maximal score among the orderings of a set of nodes $A$ where $B$ are potential parents of all nodes in $A$.

\section{Exact causal discovery}
\label{sec:exact_discovery}

Here we present a version of the exact causal discovery algorithm of \cite{koivisto2004exact}, rephrased in terms of right (left) orders that we will need later.

\begin{figure*}[t]
    \centering
\includegraphics[width=0.75\textwidth]{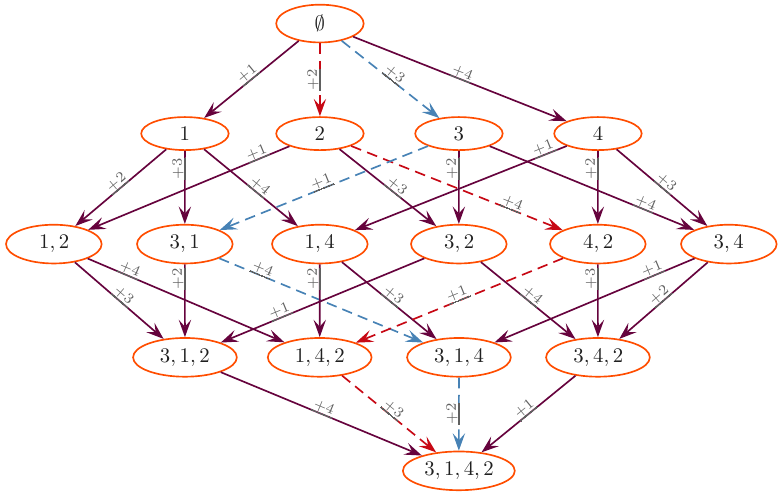}
    \caption{\textbf{The ordered power set representation of pruning all $p!$ orders to $2^p$ ordered subsets.} Moving down the diagram we append the new node indicated on the edge to the front (back) of the previous right (left) suborders. We compare the incoming suborders at each (orange) node and only keep the best. Arriving at the bottom of the diagram we have the OOO. The dashed paths arrive at the OOO when appending at the front (red) or back (blue).}
    \label{fig:exactorderposet}
\end{figure*}

\subsection{Permutation trees}

Since the score of each node only depends on potential parents, we can compute the score of each right (left) order by adding new nodes to the front (back) one at time. This creates a tree of all $p!$ orders, from which we can exhaustively find an optimal order. To reduce complexity, one can use the fact that optimality of an order is inherited by its suborders:
\begin{theorem}
\label{th:suborder}
For a given optimal right (left) order $\vr{n}$ ($\vl{n}$) of $U \subset V$, the right (left) suborder $\vr{n-1}$ ($\vl{n-1}$) is an optimal right (left) suborder of $U\setminus \{\o_{n}\}$.
\end{theorem}
\begin{proof}
Suppose $\vr{n-1}$ is not an optimal right suborder of $U\setminus \{\o_{n}\}$. Then there exists a suborder $\vrp{n-1}$ with $S(\vrp{n-1}) > S(\vr{n-1})$. Appending $\o_{n}$ in front
\begin{align*}
& S(\ro{\o_{n},\vnp_{n-1:1}}) = \s{\o_n}{\vnp_{n-1:1}} + S(\vrp{n-1})  > \s{\o_n}{\vnp_{n-1:1}} + S(\vr{n-1}) = S(\vr{n}) 
\end{align*}
then $\vr{n}$ is not optimal, which is a contradiction. The proof for left suborders is analogous.\end{proof}
\begin{corollary}
For a given optimal right (left) suborder $\vr{n}$ ($\vl{n}$) of $U \subset V$, the right (left) suborder $\vr{n-k}$ ($\vl{n-k}$) for $1\leq k\leq n$ is an optimal right (left) suborder of $U \setminus \{\o_j\}_{j=n-k+1}^n$ .
\end{corollary}
\begin{proof}
Apply \tref{th:suborder} several times.
\end{proof}

\subsection{Ordered power set}

To reduce the complexity from factorial, \cite{koivisto2004exact} essentially used Theorem \ref{th:suborder} to prune suborders with the same set of nodes by comparing their scores and keeping only one with the highest score. We illustrate this in \fref{fig:exactorderposet} where we project the permutation tree onto the ordered power set or Hasse diagram of the node subsets. 

At stage $n$, we have the $\binom{p}{n-1}$ orders from stage $(n-1)$ each appended with $(p-n+1)$ new elements to create $(p-n+1)\binom{p}{n-1} = p\binom{p-1}{n-1} = n\binom{p}{n} $ suborders of length $n$.
\begin{pruning}[Suborder pruning]\label{prune:suborder}
At stage $n$, we go through each group of $n$ suborders with the same set of nodes and keep only one with the highest score. If several suborders have the same score, we retain the one with the last two nodes in numerical order (lower node number to the left).
\end{pruning}
\begin{theorem}
\label{th:OOO}
The OOO will not be pruned.
\end{theorem}
\begin{proof}
By \tref{th:suborder}, any suborder of the OOO will be amongst the highest scoring on that set of nodes, while the transposable nodes of the OOO are in numerical order by construction.
\end{proof}
\begin{corollary}
The resulting order at the last stage of the tree pruning is the OOO.
\end{corollary}
This is, for example, the ordered subset at the bottom of the power set of \fref{fig:exactorderposet}.

\subsection{Complexity} \label{sec:old_complexity}

In the setting of \cite{koivisto2004exact}, since the power set is complete we can compare the scores of all the $n\binom{p}{n}$ suborders at stage $n$ in linear time using their indexing in the Hasse diagram, giving a total complexity of $\bigO(p2^p)$ for the tree pruning. In addition, the lookup of each score when appending a new node is assumed to be $\bigO(1)$, so scoring all the $n\binom{p}{n}$ new suborders at stage $n$ also leads to a total complexity of $\bigO(p2^p)$.
The alternative approach of \cite{silander2006simple} which exhaustively checks for the best sinks to build up an OO from the front likewise has complexity of $\bigO(p2^p)$. The problem can also be recast as a reinforcement learning task \citep{mokhtarian2023novel}, though this also does not reduce the complexity.

\cite{koivisto2004exact} showed how to use the M\"obius transform to reduce the complexity to $\bigO(2^p)$, however this relies on precomputing various score tables. They allowed each node to have up to $K$ possible parents from any of the $p$ available and computing the score table takes $\bigO(C(m, p, K)p^{K+1})$, where $C$ is the time of each local score computation, possibly depending on the data size $m$. Thereafter scores are combined giving an overall complexity of $\bigO(p2^p)$.

In our setting we allow $K$ parents from a preselected set so that creating the original score tables is $\bigO(C(m, p, K)p2^{K})$ and computing combined score tables is $\bigO(p2^{K})$ \citep{kuipers2022efficient}. In order to not fully trust the preselected search space, we allow one additional parent outside. Appending a new node to a suborder involves looking up $\bigO(p)$ scores. After the score tables have been created, running through the power set and scoring the suborders leads to an overall complexity of $\bigO(p^22^p)$, which will be the baseline to compare to when we start pruning.

\section{Lowering the complexity of exact causal discovery}
\label{sec:pruning}

In the following we show that we can build the full optimal order starting with the empty order, and that at each step $n$ we can prune the suborders of length $n$ that will not lead to the optimal order. In particular, when we add nodes we only need to keep those whose placement is optimal.
\begin{theorem}
\label{th:contain}
Let $M_0=([\dots])$ and let for each $n\le p$, $M_n$ be the set of right (left) orders of length $n$ created by adding a node at the front (back), only if that node is optimal there, of some order in $M_{n-1}$.
Then each $M_n$ contains the suborder of an optimal order  $\vn^* \in M_p$.
\end{theorem}
\begin{proof}
We prove inductively that each $M_n$ contains the suborder of an optimal order $\vn^*\in M_p$.
For $n=1$, every node is optimal at the back so $M_1=\{\ro{j}\}_{j=1}^p$ and it is clear that some element here is a suborder of the optimal order.
For $n>1$ we assume inductively that  $M_{n-1}$ contains the suborder of the optimal full order.
Let $\vr{n}$ be a suborder of $\vn^*$. 
Then we need to prove that it can be constructed from some order in $M_{n-1}$.
Clearly, $\vr{n}$ is created from $\vr{n-1}$ by adding $\o_{n}$ at the back.
Since $\vr{n}$ is a suborder of $\vn^*$ we know by Corollary 1 that it is optimal on $\{\o_j\}_{j=1}^n$.
By Theorem 1, $\vr{n-1}$ is optimal on $\{\o_j\}_{j=1}^{n-1}$ so by the induction hypothesis it must be contained in $M_{n-1}$.
\end{proof}
\begin{corollary}
A full OO can be built by only adding new nodes to the front (back), starting from the empty order.
\end{corollary}
\begin{proof}
By \tref{th:contain}, $M_p$ will contain the optimal order.
\end{proof}

\subsection{Pruning right orders}

We focus on pruning specifically for right suborders.
If a node added to the front of a right order were to fit better further along the order, then we can prune by \tref{th:contain}.
\begin{pruning}[Optimal front] \label{prune:opt_front}
We prune $\vr{n}$ if $\osc{\ro{\vn_{n-1:k},\vn_{n},\vn_{k-1:1}}} > \osc{\vr{n}}$ for some $k < n$.
\end{pruning}
We can further ensure numerical ordering in the case of equal score under transposition. 
\begin{pruning}[Ordered front] \label{prune:ord_front}
We prune $\vr{n}$ if $\vn_n>\vn_{n-1}$ and $\osc{\vr{n}} = \osc{\ro{\vn_{n-1},\vn_{n},\vn_{n-2:1}}}$.
\end{pruning}
where $>$ is the usual mathematical operator since the vertex set just consists of the numbers $1,\ldots,p$.
Both these pruning rules have been employed previously \citep{van2015machine}. For further pruning, however, we use the following observation:
\begin{lemma} \label{lemma:dormant}
For any partition $a\cup b$ of the dormant nodes in a right suborder $\vr{n}$ it holds that
\begin{align}
\label{eq:ineq}
 & \smax{a \cup b}{\vn_{n:1}} \le  \smax{a}{b, \vn_{n:1}}+ \smax{b}{a , \vn_{n:1}}.
\end{align} 
\end{lemma}
\begin{proof}
On the right the nodes in $a$ and $b$ are allowed all the nodes in $b$ and $a$ as parents respectively, while on the left there are restrictions induced by the ordering of $a \cup b$, so those scores can only be lower or equal.
\end{proof}
\begin{theorem} \label{th:nogaps}
If 
$\osc{\vr{n}} = \osc{\lro{\vn_n}{\vn_{n-1:1}}}$
then  $\underset{\o \in \ro{ \vn_{n-1:1}}}{\mathrm{\max}}\, \osc{\o} \le \underset{\o \in \ro{ \vn_{n:1}}}{\mathrm{\max }}\, \osc{\o}$.
\end{theorem}
\begin{proof}
We denote by $\ghvec = V \setminus \{ \o_j\}_{j=1}^{n}$ the dormant nodes of $\vr{n}$. The maximal score of an order in $\vr{n-1}$  is 
\begin{align*}
& \underset{\o \in \ro{\vn_{n-1:1}}}{\mathrm{\max }}\, \osc{\o} = \smax{\ghvec,\vn_n}{\vn_{n-1:1}} + \osc{\vr{n-1}} \\
& \le \smax{\ghvec}{\vn_{n:1}} + \s{\vn_n}{\ghvec,\vn_{n-1:1}} + \osc{\vr{n-1}} \text{ / by \eqref{eq:ineq} } \\
& =   \smax{\ghvec}{\vn_{n:1}} + \osc{ \lro{\vn_n}{\vn_{n-1:1}}  }  =  \smax{\ghvec}{\vn_{n:1}} + \osc{\vr{n}} = \underset{\o \in \ro{\vn_{n:1}}}{\mathrm{\max }}\, \osc{\o}.
\end{align*}
\end{proof}
\tref{th:nogaps} means that if the score of $\vn_n$ is independent of the dormant nodes, the total order score cannot be improved by putting some other node at the front. Since we have a numerical ordering constraint in the case of equal scores, we can prune for any node with a lower number and only keep particles with the highest number node in front whose score is independent of the dormant nodes.
\begin{pruning}[No right gaps] 
\label{prune:norightgaps}
Consider all the right orders of length $n$ created from $\vr{n-1}$.
Let $\vn_{n}'$ be the maximal node number such that $\osc{\ro{\vn_{n}',\vn_{n-1:1}} } = \osc{\lro{\vn_{n}'}{\vn_{n-1:1}}}$.
We prune all suborders $\vr{n}$ where $\vn_{n} < \vn_{n}'$.
\end{pruning}

In order to prune further, we consider the dormant nodes in more detail.
\begin{theorem} \label{th:rightdormant}
Let $\vr{n} \in M_n$ and $\vn^*$ be an optimal order of $V$. 
Then $\vn^* \notin \vr{n}$ if for any dormant node $\ghnode$ and $k\le n$ we have 
$\osc{\ro{\vn_{n:k},\ghnode,\vn_{k-1:1}}} > \osc{\lro{\ghnode}{\vn_{n:1}}}$.
\end{theorem}
\begin{proof}
With $\ghnode$ a single dormant node and $\ghvec = V \setminus h \cup \{ \o_j\}_{j=1}^{n}$ the set of other dormant nodes, we show that the maximal score of an order in $\ro{\vn_{n},\ghnode,\vn_{n-1:1}}[\ghvec]$ is greater than the maximal score in $\ro{\vn_{n:1}}[\ghnode,\ghvec]$. 
From the setting we have
\begin{align}
\label{eq:setting}
\osc{\lro{\ghnode}{\vn_{n:1}}} &= \osc{\lo{\ghnode}} + \osc{\vr{n}} < \osc{\ro{\vn_{n-k},\ghnode,\vn_{k-1:1}}}.
\end{align}
The maximal score of an order in $\vr{n}$  is 
\begin{align*}
&  \underset{v \in \ro{\vn_{n:1}}}{\mathrm{\max }}\, \osc{v} =\smax{\ghvec,\ghnode}{\vn_{n:1}} + \osc{\vr{n}}  \le \s{\ghnode}{\ghvec,\vn_{n:1}}+ \smax{\ghvec}{\ghnode, \vn_{n:1}} + \osc{\vr{n}} \text{ / by \eqref{eq:ineq} } \\
& =   \smax{\ghvec}{\ghnode, \vn_{n:1}} + \osc{ \lro{\ghnode}{\vn_{n:1}}  }  <  \smax{\ghvec}{\ghnode,\vn_{n:1}} + \osc{\ro{\vn_{n:k}, \ghnode, \vn_{k-1:1}}} \text{ / by \eqref{eq:setting}} \\
& = \underset{v \in \ro{\vn_{n:k},\ghnode,\vn_{k-1:1}}}{\mathrm{\max }}\, \osc{v},
\end{align*}
which is the maximal score of an order in $\ro{\vn_{n:k},\ghnode,\vn_{k-1:1}}$, so $\o^* \notin \vr{n}$.
\end{proof}
\begin{pruning}[Removing dormant right gaps]\label{prune:dormantrightgap}
We prune $\vr{n}$ if for any dormant node $\ghnode$, and for some $k\le n$, we have $\osc{\ro{\vn_{n:k},\ghnode,\vn_{k-1:1}}} > \osc{\lro{\ghnode}{\vn_{n:1}}}$.
\end{pruning}
In other words, we check if $\vn_n$ is pushed as far to the right as possible, or if the total score could be improved by inserting another node ($h$) as a potential parent. If so, we prune. If they have equal score, we impose the following sorting criteria.
\begin{pruning}[Ordering dormant right gaps]\label{prune:dormantrightord}
We prune $\vr{n}$ if there is a dormant node $h>\vn_n$ such that $\osc{\ro{\vn_{n},\ghnode,\vn_{n-1:1}}} = \osc{\lro{\ghnode}{\vn_{n:1}}}$.
\end{pruning}

\subsection{Global pruning}

Along with the pruning rules for right (left) orders, we have the global suborder pruning (Pruning 1) which we restate for potentially differing numbers of suborders:
\begin{pruning}[Removing duplicates] \label{prune:no_dup}
At stage $n$, we have $N$ orders, $M_n^N$, of length $n$. We go through all the orders with the same set of nodes and keep only one with the highest score, or in numerical order with the same score.
\end{pruning}

One can also consider left orders of length $n$ as subject for pruning, which we discuss in the Supplementary Material (Section \ref{sec:left}), where we also discuss path-search pruning (Section \ref{sec:path}) and detail the computational complexity of the scheme (Section \ref{sec:complex}). We outline the main order pruning approach in Algorithm \ref{alg:main}, while the pseudocode for the detailed steps and each Pruning is detailed in the Supplementary Material (Section \ref{sec:algos}).

\begin{algorithm}[t]
\DontPrintSemicolon

\hrulefill
\caption{Main algorithm}\label{alg:main}
\vspace{-1.5ex}
\hrulefill

$M_{p} \gets \{\}$ \; 

\For{$n = 1$ \KwTo $p$}{ % For loop starts here

    $M_{n} \gets \{\}$\;
    
    \eIf{$n = 1$}{ % If condition for n = 1
        \For{$j \in V$}{ % Inner For loop for j in V
            $\bar{m} \gets \ro{j}$\;
            
            \If{$hasGap(\bar{m})$}{ % If condition to check gap
                \textbf{continue}\; % Continue statement in bold
            }
 
            Update insertion scores for $\bar{m}$\; \tcc*[r]{$\bigO(p^2)$}
            $M_n \gets M_n \cup \{\bar{m}\}$\;
        } 
        % End of inner For loop
    }{ 
    % End of If for n = 1 
    % Else condition for mainTransition
        $M_{n} \gets mainTransition(M_{n-1})$\;
    }
} % End of outer For loop
\Return $M_p$\;

\vspace{-1.5ex}
\hrulefill
\end{algorithm}

\subsection{Divide-and-conquer}

The result for matchings (Section \ref{sec:matching}) suggests that if the best network is disconnected we could leverage this to further speed up the causal discovery. To extend pairwise disconnection to general networks we first consider how each parent affects the scores for each child.

We define matrices $H$, whose entries for node $\vn_i$ are the biggest increase in score from including $\vn_j$ as a parent of $\vn_i$, compared to excluding it, for any possible parent set:
\begin{align}
H^{\max}_{i,j} = & \max_{\Pa_{\vn_i}} \left\{ \sigma(\vn_i,\Pa_{\vn_i} \cup \vn_j \mid D) -  \sigma(\vn_i, \Pa_{\vn_i} \setminus \vn_j \mid D) \right\}
\end{align}
%
% decreases 3,1,2,-2,5,-1: largest = 3
% increases -3,-1,-2,2,5,1: smallest = -3
and the biggest decrease in score (most negative change) when including $\vn_j$ as a parent of $\vn_i$:
\begin{align}
H^{\min}_{i,j} = & \min_{\Pa_{\vn_i}} \left\{\sigma(\vn_i,\Pa_{\vn_i} \cup \vn_j \mid D) -  \sigma(\vn_i, \Pa_{\vn_i} \setminus \vn_j \mid D) \right\}
\end{align}
The next theorem states that if all parent sets of any node $\vn_i \in A\subset V$ get higher (or equal) scores when excluding any node $\vn_j \in B\subset V$, and vice versa, then $A$ and $B$ must be disconnected in the optimal DAG. 
\begin{theorem} \label{th:disconnect1}
\emph{If} for any subsets $A,B \subset V$ we have $H^{\max}_{A,B} \leq 0$ and $H^{\max}_{B,A} \leq 0$,
then $A,B$ are disconnected in an optimal DAG.
\end{theorem}
\begin{proof}
Assume there is an edge from $a\in A$ to $b \in B$ in the optimal DAG, $\DAG$. Since $H^{\max}_{b,a} \leq 0$, removing that edge would not decrease the score, so either $\DAG$ was not optimal which is a contradiction, or there is another optimal network without the connection.
Likewise, if there is an edge from $b \in B$ to $a\in A$.
\end{proof}
\begin{corollary} \label{divcon:upper}
For disconnected components, denoted $\mathcal C_u$, of the graph induced by $H^{\max} > 0$, we can run the order search independently.
\end{corollary}
This allows us to separate the network into sets of corresponding random variables which are independent.
However, it may not find all disconnected components in the optimal network, so to proceed we also consider $H ^{\min}$.
If $A\subset V$ and $B \subset V$ are in different components of an optimal DAG, we know that adding any edge from $b \in B$ to $a \in A$ would reduce (or leave equal) the score of the parent set of $a$, so $H^{\min}_{a,b} \leq 0$. Likewise for edges from $A$ to $B$. This argument leads to:
\begin{theorem} \label{th:disconnect2}
\emph{If} subsets $A,B \subset V$ are disconnected in an optimal DAG \emph{then} $
\forall a\in A, b\in B$, we have $H^{\min}_{a,b} \leq 0$ and $H^{\min}_{b,a} \leq 0$.
\end{theorem}
Disconnected subcomponents from the matrix $H^{\min}$ might not be disconnected in the optimal network. Instead
\begin{corollary} \label{divcon:lower}
Disconnected subcomponents, denoted $\mathcal C_l$, in the graph induced by the boolean matrix $H^{\min} > 0$ are potentially disconnected in an optimal DAG.
\end{corollary}
We are now able to formulate the following divide-and-conquer algorithm to find the optimal DAG.

\begin{divcon*}
$\left.\right.$
Each subcomponent $\mathcal{C}_l$ from Corollary \ref{divcon:lower} is contained in one component of $\mathcal{C}_u$ from Corollary \ref{divcon:upper}, which can be treated independently. Therefore we first organize them accordingly and then find the optimal DAG as follows:

\begin{enumerate}
    \item Run the order search independently for each component $A \in \mathcal{C}_l$ with all other nodes as potential parents. Let this optimal ordering of $A$ be denoted by $\harpoonl{v}_A$ and the corresponding optimal DAG by $\DAG_{A|V\setminus A}$.
    \item Treat the components $\mathcal{C}_l$ as the node set of a network $\N$. For any pair of nodes $A, B$ in $\N$, draw an edge $A\leftarrow B$ if any node in $\DAG_{A|V\setminus A}$ has a parent in $B$.
    \item Find the cycles in $\N$. Merge nodes along cycles and intersecting cycles to form a new component set $\mathcal{C}_l$. If no cycles were found go to step 4, otherwise return to step 1 and rerun the order search for modified components.
    \item An OO is obtained by concatenating each optimal order $\harpoonl{v}_A$ corresponding to the nodes $A$ in $\N$ according to a topological order of $\N$. The union of the DAGs $\DAG_{A|V\setminus A}$ gives the corresponding optimal DAG.
\end{enumerate}
\end{divcon*}

\section{Polynomial causal discovery}\label{sec:poly}

Here we use components of the pruning rules above to prove polynomial exact causal discovery for certain classes of causal Bayesian networks. We note that this is distinct from previous results which find the optimal network within a restricted class \citep{chow1968approximating,gaspers2015finding, gruttemeier2022learning} in polynomial time but do not prove the network is optimal overall. Instead, we are guaranteed to find the optimal network in our exact discovery and we may do so in polynomial time if the optimum satisfies certain criteria.

For example, when the empty graph is optimal, we trivially have that each order has the same score, and the score of each node does not depend on the placement of the others. Since we use the ordering restrictions, for example in Pruning~\ref{prune:ord_front} and \ref{prune:dormantrightord}, then we would prune every order apart from the numerically ordered one and have a single particle at each iteration. Including the computational complexity of each iteration (as detailed in Section \ref{sec:complex}), overall this leads to a cubic algorithm, once we have the score tables. This result is however subsumed by our polynomial proof for matchings (detailed in Section \ref{sec:matching}) and the following. 

For sparse networks, our divide-and-conquer approach to finding disconnected components will lower the complexity to that of the largest component. For example, if we consider graphs with a maximum number of edges $E$, the largest component is bound by $E$.
Since the number of components is bound by $p$, the number of merging potential components and re-running the optimal search is also at most $p$, following the divide-and-conquer algorithm. For each component, finding an optimal network is exponential in $E$ in the worst case, but this does not depend on $p$, and hence is not exponential in $p$. Overall, the exact causal discovery remains polynomial in $p$.

More generally, for many random sparse networks, like Erd\"os–R\'enyi graphs with a density $\lesssim 1$, there is a high probability there is no giant component and that the largest components are of order $\log(p)$ \citep{molloy1995critical,newman2001random}. Finding the components involves at most linear (in $p$) network searches, while each search is now exponential in $\log(p)$, and hence polynomial in $p$. Therefore, for any sparse network, as long as the largest connected component remains logarithmic in $p$ we have a polynomial algorithm for exact causal discovery.

These subsets of causal Bayesian networks then constitute an exception from the general NP-hardness of exact causal discovery \citep{chickering1996learning,chickering2004large}.

\section{Simulation studies} \label{sec:sims}

To examine the computational complexity in practice, we performed simulation studies for networks of sizes $p \in \{10,\dots,30\}$ with a sample size of $300$. 
For each network size, random graphs were generated using the default \texttt{randDAG} function from the \texttt{R}-package \textbf{pcalg} \citep{kalisch2012causal} with average density (neighbourhood size) of $d\in\{0,0.1,\dots, 2\}$.
Continuous data were generated from a linear Gaussian structural equation model, where strengths of the edges (regression coefficients) were sampled uniformly in the range [0.25, 1]. % and standardised. 
This process was repeated with 1000 different seeds.

As a preliminary step to run our algorithm, we first calculated the score tables using the BGe score \citep{geiger2002parameter,kuipers2014addendum} with mean hyper-parameter $\alpha_\mu=0.1$.
We estimated a DAG search space using an iterative MCMC procedure \citep{kuipers2022efficient}, which includes the skeleton from constraint-based testing and of the estimated DAG, and where each node is allowed an additional parent outside that core set. The procedure used the \texttt{R}-package \textbf{BiDAG} \citep{suter2023bayesian} and exported the output to our \texttt{C++} program (available at \url{https://github.com/felixleopoldo/dncDagger}).
To track the complexity, we record the number of suborders remaining after pruning, $N_n$, at each stage of the main algorithm (Algorithm \ref{alg:main}) and the total runtime. We also compute the total number of suborders $\Sigma_N = \sum_{n} N_{n}$. Since this total is bound by $2^p$, this is the main component where we expect exponential complexity.

\begin{figure}[t]
\centering
\begin{tabular}{cc}
\includegraphics[width=0.49\textwidth]{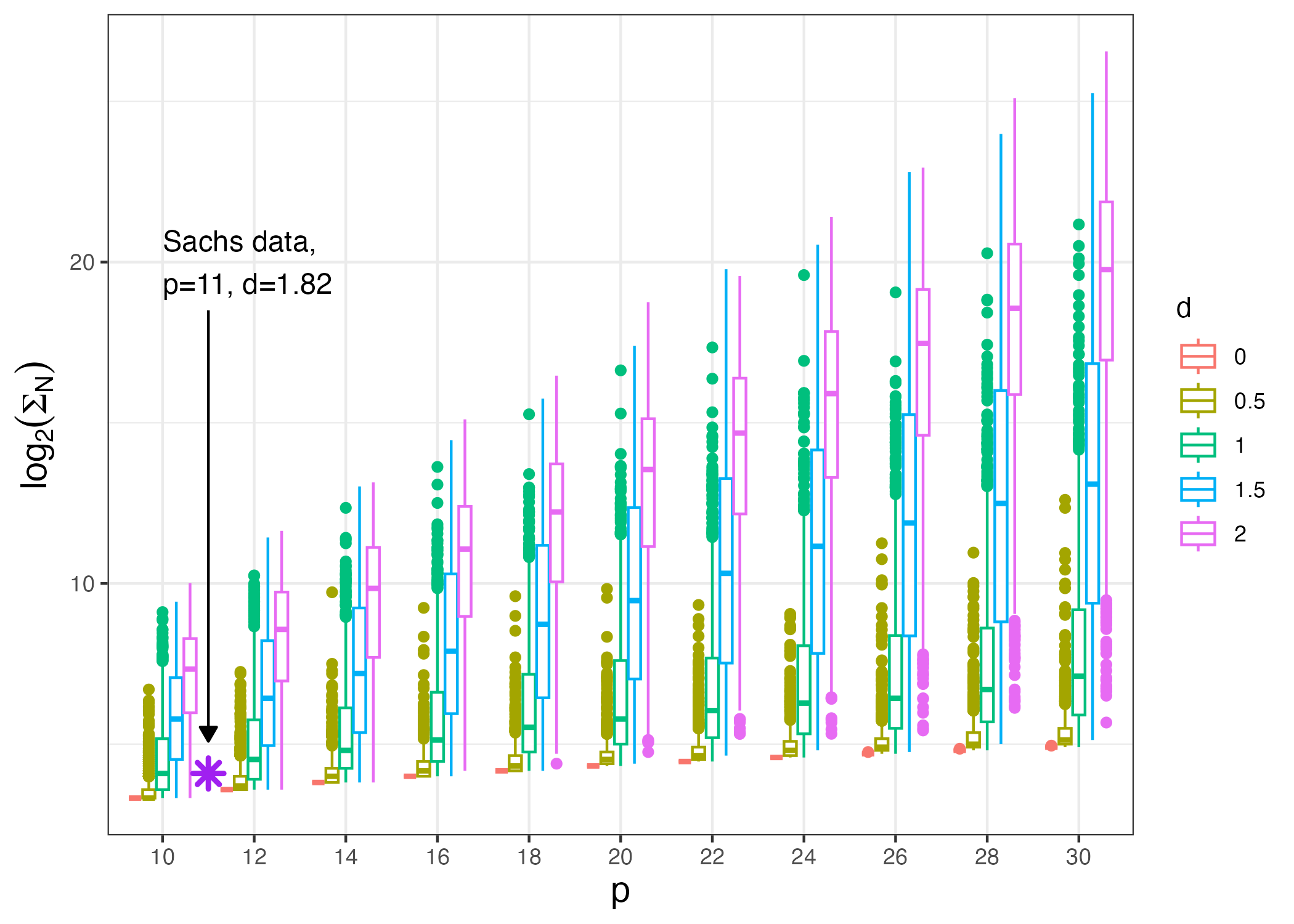}
&
\includegraphics[width=0.49\textwidth]{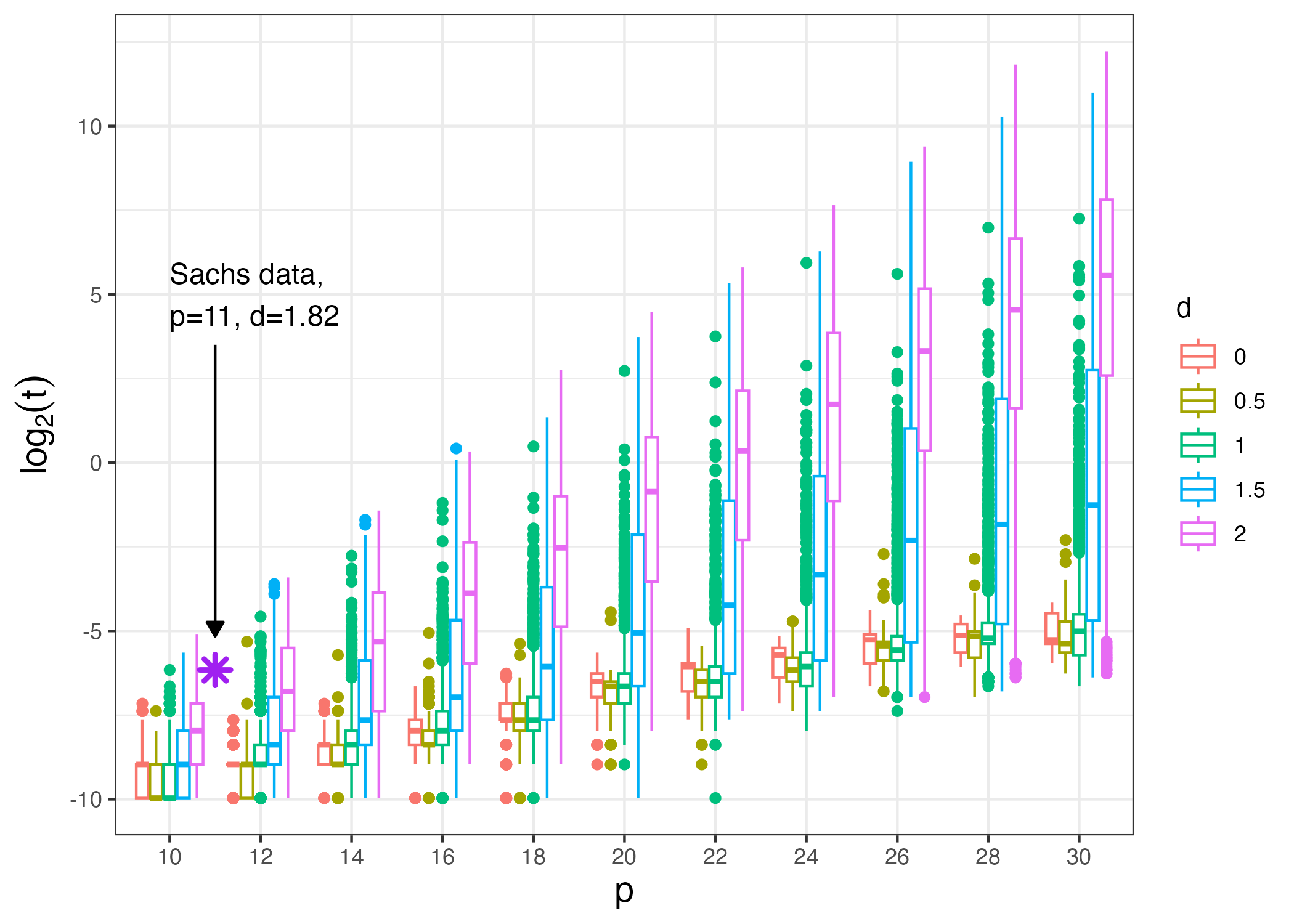}
\end{tabular}
\caption{The logarithm of the total number of suborders $\Sigma_N$ and the overall computational runtime $t$ (in seconds) as a function of the network size $p$ and stratified by the network density $d$. We additionally overlay results for the often-studied Sachs dataset \citep{sachs2005causal}.}
\label{fig:boxplots}
\end{figure}

The results (\fref{fig:boxplots} and Supplementary \fref{fig:boxplots_wide}) show up to linear behaviour in the growth of the logarithm of the total number of suborders for increasing network dimension, but also that it depends heavily on the network density. In fact for $d=0$ and no edges, when the empty graph has the highest score, our ordering constraint means that we only keep one suborder at each stage so $\Sigma_N = p$, and we get sub-linear behaviour with the logarithm. For $d=0.5$ the majority of cases also have a low number of particles, as we typically expect small components at this density, but some outliers requiring significantly more particles start to appear.  

\begin{figure*}[t]
\includegraphics[width=0.98\textwidth]{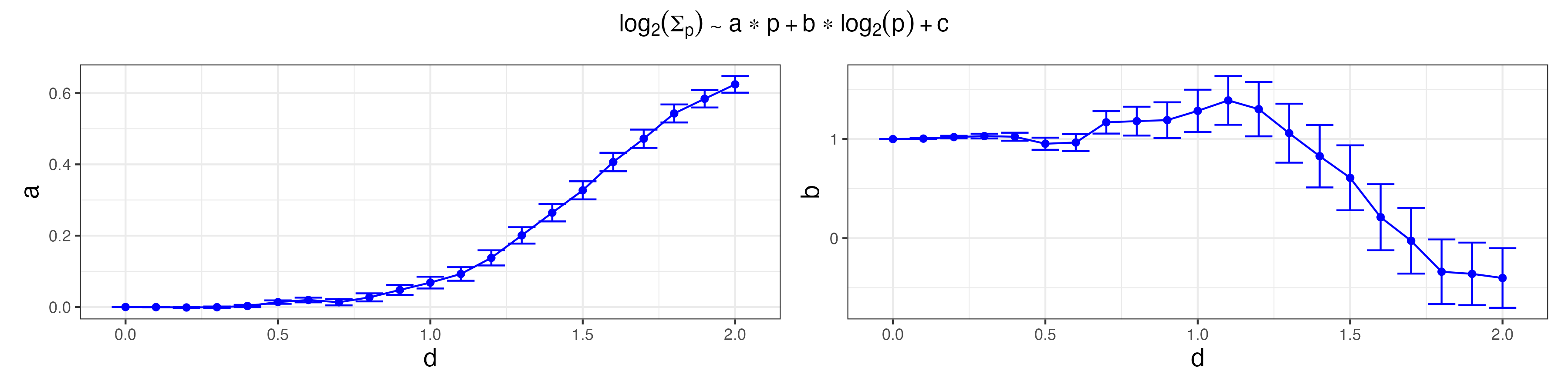}
\caption{The regression coefficients of the logarithm of the total number of suborders $\log_{2}(\Sigma_N)$ against $p$ and $\log_2(p)$, with standard errors, as a function of the network density $d$ (the expected neighbourhood size).}
\label{fig:regrplots}
\end{figure*}

To examine this more carefully, we fit a regression line for the total number of suborders: $ \log_2(\Sigma_N) = ap + b \log_2(p) + c$ for each density. The fitted coefficients (\fref{fig:regrplots}) show that we have no exponential term ($a=0$) up to a density of around 0.4 (and a linear term, $b=1$), and then the logarithm of the base of the exponential term increases slowly up to a density of around 1.2 (in line with the increasing spread of outliers in \fref{fig:boxplots}) and then more linearly for higher densities, while the polynomial term decreases to around $b=0$. For the densities up to 2 considered, the complexity remains substantially below the theoretical maximum of $a=1$ (and $b=0$, when all possible $2^p$ suborders are kept). The fact that most examples require fewer particles at lower density is clearer when we perform a median regression (Supplementary \fref{fig:regrplots_med}) with lower exponential term $a$.   

In terms of runtime (\fref{fig:boxplots}), the majority of cases even with $d=1$ complete very quickly, though some examples take much longer, and the typical time also increases at the higher densities. For the runtime per suborder (Supplementary \fref{fig:ratio_boxplot}), we would expect up to quadratic behaviour in $p$ according to the complexity analysis, the actual time seems more linear for higher densities suggesting the complexity may be sub-quadratic in practice. Taken together, the practical runtime is much lower than the worst case $\bigO(p^22^p)$. As this is the best case without pruning, our pruning rules do seem to allow us to significantly lower the complexity of exact causal discovery.

As a simple example, we consider the infamous Sachs dataset \citep{sachs2005causal} which measured the levels of 11 phosphoproteins and phospholipids in individual T-cells. We use the second control dataset, log-transformed, for which most causal discovery algorithms return the same network \citep{rios2021benchpress}. We find the optimum by only needing to track a very small number of suborders (17, annotation in \fref{fig:boxplots}), as the network is disconnected (Supplementary \fref{fig:Sachs_net}). Despite the low complexity of exact discovery in this case, the actual runtime is in line with networks of similar size and density due to overheads in the implementation, indicating room for further improvements. 

Since we are considering the task of exact causal discovery, where we can prove the result is optimal (as discussed in the Introduction), we additionally compared to the ILP approach of GOBNILP \citep{bartlett2017integer,cussens2017bayesian, cussens2020gobnilp} by passing it the tables of scores from the same search space. Though GOBNILP uses highly optimised solvers, it only has a runtime advantage at the higher density (Supplementary \fref{fig:vsGOBNILP}), while in line with \fref{fig:regrplots}, our polynomial and then lower exponential complexity at lower densities provides us with better runtimes for sparser networks.

\section{Discussion}\label{sec:discussion}

We present an algorithm for exact causal discovery which improves on dynamic programming methods \citep{koivisto2004exact, silander2006simple} by pruning orders which cannot possibly lead to the global optimum, and which expands upon previous pruning approaches \citep{van2015machine}. We also include pruning rules similar to the A$^{*}$ search approach of \cite{yuan2013learning}, which finds the shortest (least costly in terms of negative score) path through the ordered power set of node orders (\fref{fig:exactorderposet}) and prunes the power set by removing paths that cannot possibly improve the score. While they rely on various heuristics to prune, based on scores with potential cycles and then on avoiding cycles, we focussed on bounds at lower computational cost. When the optimal solution is a matching, our algorithm only needs to keep track of one suborder at each stage and can exit immediately, so the total complexity is polynomial. While finding the best matching amongst matchings is well-known to be polynomial, as is finding the best tree amongst trees \citep{chow1968approximating}, our study instead targets the exact discovery of finding the best network amongst all DAGs without restrictions on the type of network. Our algorithm not only completes in polynomial time when the optimum is a matching, but additionally proves the global optimality of the solution amongst all networks. We further extended our approach with a novel divide-and-conquer algorithm to detect disconnected components in the optimal DAG in an effective manner. For sparse networks with a logarithmic largest component, we can also prove optimality in polynomial time. These results constitute a remarkable exception to the general NP-hardness of causal discovery, even with a search space with limited in-degree \citep{chickering1996learning,chickering2004large}.

We design our pruning rules with no complexity overhead so that our worst-case complexity is the current best case (without pruning). Simulation studies show that the actual runtime and complexity can be much lower. In particular, beyond the polynomial behaviour for sparse networks, we find a slowly and then roughly linearly increasing trend between the logarithm of the base of the exponential term in the complexity and the density of the simulated networks. In our simulations, we start seeing some outliers with exponential runtime at a density of $0.5$. Aymptotically we expect a sharp phase transition for random graphs with almost surely no giant component below a critical value (a density of 1 in our case), and almost certainly a giant component above. Likewise, we would expect a sharp transition from polynomial to exponential complexity asymptotically at this critical value. In practice, and away from asymptotics, our approach scales well for larger sparser networks.

We worked on the search space used by an iterative hybrid scheme \citep{kuipers2022efficient} where each node may have parents from within preselected sets, plus an additional one outside that set. More typical for sampling had been to consider all possible parent sets up to a maximum in-degree \citep{friedman2003being,kuipers2017partition}, or for discovery to prune further to tables of candidate parent sets \citep{friedman1999learning,teyssier2005ordering}. This is the standard input to current state-of-the-art exact causal discovery algorithms \citep{malone2018empirical}. The hybrid setting offers several advantages. First, it typically offers good performance in causal discovery \citep{rios2021benchpress}. Second, if the initial search space is found with a consistent method, like the pc-algorithm, and if the score function is consistent, like with a penalised likelihood, then we retain consistency in our exact causal discovery. Third, the hybrid approach allows us to precompute all scores needed for the search, rather than potentially needing to look through tables of scores of candidate parent sets. As such we can remove this factor from influencing the algorithm runtime. In particular, we observe clear trends of the runtime as a function of the density of the network and its size, which seems much more predictable than for other exact solvers \citep{malone2018empirical}. We can therefore examine the hardness of exact causal discovery more precisely. 

At lower densities, our order-based pruning algorithm is generally faster than the ILP approach of GOBNILP
\citep{bartlett2017integer,cussens2017bayesian, cussens2020gobnilp}, which was also typically found to be the fastest compared to other exact algorithms \citep{malone2018empirical}, especially for smaller and sparser instances \citep{trosser2021improved}. Our new divide-and-conquer and pruning rules might help improve ILP approaches, while these may suggest other avenues for better pruning our search. Overall, our approach offers a useful framework for further lowering the complexity of exact causal discovery and extending its use for larger and sparser networks. 

%\clearpage

\bibliography{exactcausal}

%%%%%%%%%%%%%%%%%%%%%%%%%%%%%%%%%%%%%%%%%%%%%%%%%%%%%%%%%%%%

%\clearpage

\appendix

\renewcommand{\figurename}{Supplementary Figure}
\setcounter{section}{18}
\setcounter{figure}{0}
\renewcommand{\thesection}{\Alph{section}}

\clearpage

\section{Supplementary Material}

\subsection{Pruning left orders} \label{sec:left}

Here we consider left orders of length $n$ as subject for pruning. The simplest strategy is again when the last node added could fit better earlier in the order.
\begin{pruning}[Optimal back] \label{prune:opt_back}
We prune $\vl{n}$ if for some $k < n$ we have $\osc{\lo{\vn_{1:k-1},\vn_{n},\vn_{k:n-1}}} > \osc{\vl{n}}$.
\end{pruning}
Analogously to Pruning~\ref{prune:ord_front},
we ensure the two nodes in the back are in numerical order if they can be swapped without affecting the score.
\begin{pruning}[Ordered back] \label{prune:ord_back}
We prune $\vl{n}$ if $\vn_n<\vn_{n-1}$ and $\osc{\vl{n}} = \osc{\lo{\vn_{1:n-2},\vn_{n},\vn_{n-1}}}$.
\end{pruning}
We also consider the analogue of \tref{th:nogaps} 
for left orders when appending new nodes to the back.
\begin{theorem} \label{th:leftone}
If $\osc{\vl{n}} = \osc{\lro{\vn_{1:n-1}}{\vn_n}}$, then $ 
 \underset{\o \in \lo{\vn_{1:n-1},\ghnode}}{\mathrm{\max }}\, \osc{\o}
 \ge
 \underset{\o \in \lo{\vn_{1:n},\ghnode}}{\mathrm{\max }}\, \osc{\o}
 $ for any dormant node $\ghnode$.
\end{theorem}
\begin{proof} The equality means that the score of $\vn_n$ is independent of the other nodes. However, the score of some of the dormant nodes $\ghnode$ may depend on $\vn_n$.
In particular for any $\ghnode$ we have that
\begin{align*}
\underset{\o \in \lo{\vn_{1:n},\ghnode}}{\mathrm{\max }}\, \osc{\o}
\le &
\underset{\o \in \lo{\vn_{1:n-1},\ghnode,\vn_n}}{\mathrm{\max }}\, \osc{\o} \le 
\underset{\o \in \lo{\vn_{1:n-1},\ghnode}}{\mathrm{\max }}\, \osc{\o}.
\end{align*}
\end{proof}
\tref{th:leftone} implies that, with equal scores, as long as there is a dormant node with a lower number than $\vn_{n}$ then it is not necessary to keep $\vn_{n}$ at stage $n$, as it could be added later equally well or better.
\begin{pruning}[No left gaps] \label{prune:noleftgaps}
We prune $\vl{n}$ if $\osc{\vl{n}} = \osc{\lro{\vn_{1:n-1}}{\vn_n}}$ and for any dormant node $\ghnode$ we have $h< \vn_n$.
\end{pruning}
\tref{th:leftone} is the analogue of \tref{th:nogaps} 
and allows us to prune for the left orders similarly to the right orders in Pruning~\ref{prune:norightgaps}.
Since scores depend on possible parents, however, the pruning conditions differ due to this asymmetry, and the analogue of \tref{th:rightdormant}
does not seem to immediately lead to another pruning rule.

\subsection{Path-search pruning} \label{sec:path}

In the A$^{*}$ search algorithm of \cite{yuan2013learning}, suborders are pruned when an upper bound of their scores including the dormant nodes is less than (or equal) to the current best order score. We will denote the current best (full) order $\vstar$ with score $\Sstar = \osc{\vstar}$, and $\gbound{\vr{n}}$ as the upper bound score of the dormant nodes of $\vr{n}$.
\begin{pruning}[A$^{*}$]\label{prune:Astar}
We prune $\vr{n}$ if $\osc{\vr{n}} + \gbound{\vr{n}} \leq \Sstar$.
\end{pruning}
The task then is to find as low an upper bound as possible in low computational time. To proceed we use the observation (Lemma~\ref{lemma:dormant} and \cite{yuan2013learning}) that for any partition $\Lambda$ of the dormant nodes $U = V \setminus \{\vn_j\}_{j=1}^n$ in a right suborder $\vr{n}$ we have
\begin{equation}
\smax{U}{V \setminus U} \leq \sum_{i=1}^{m} \smax{\Lambda_i}{V \setminus \Lambda_i}   
\end{equation}
where the $\Lambda_i$ are the $m$ parts of $\Lambda$. In \cite{yuan2013learning} they considered separating into sets of size $k$ and finding the optimal sets that minimise the right hand side above. For $k=2$ this is equivalent to finding optimal matchings, which can be achieved in cubic time with a variant of Edmond's blossom algorithm. However, for $k>2$ the problem is computationally hard and in \cite{yuan2013learning} they focussed on heuristic approaches to find bounds.

Along with pruning, we can also update our current best order $\vstar$ if we can find a better one. Here, guided by the complexity, we follow a different heuristic approach for the upper bound, that will later allow us to prove optimality in polynomial time for matchings. We consider the best scoring tree on the dormant nodes, which we denote $\T{\vr{n}}$. To obtain this tree we define the matrix $F$ with entries
\begin{align} \label{eq:Feqn}
F_{ij} = \osc{\ro{\vn_i,\vn_j}} - \osc{\ro{\vn_i}} - \osc{\ro{\vn_j}}
\end{align}
which is the difference in score between allowing an edge between nodes $\vn_i$ and $\vn_j$ and excluding such an edge. The matrix $F$ is symmetric, since pairwise edges are score-equivalent, and the minimum value will be 0 in the case excluding the edge fits the data better.

The highest score of a tree network on the dormant nodes $U = V \setminus \{\vn_j\}_{j=1}^n$ of a right order $\vr{n}$ is
\begin{align} \label{eq:fbound}
\fbound{\vr{n}} = \sum_{e \in \T{\vr{n}}}F_{e} + \sum_{u\in U}\osc{\ro{u}} 
\end{align}
where $e$ are the edges in the undirected maximum spanning tree $\T{\vr{n}}$ found from the relevant submatrix of $F$, for example, with Prim's algorithm \citep{prim1957shortest}. Since trees are score-equivalent any direction imposed on the tree and, more importantly, any ordering of the dormant nodes compatible with the best scoring tree will have an equal or higher score than $\fbound{\vr{n}}$.

\begin{update}\label{update:treemax}
We update $\vstar$ if for any right order $\vr{n}$
\begin{align}
\fbound{\vr{n}} + \osc{\ro{\vr{n}}} > \Sstar
\end{align}
\end{update}

In addition we can also compare to the current best order 

\begin{update}\label{update:vstar}
We update $\vstar$ if for any right order $\vr{n}$
\begin{align}
\osc{\fo{\vstar \setminus \vr{n},\vr{n}}} > \Sstar
\end{align}
where $\vstar \setminus \vr{n}$ denotes removing the elements of $\vr{n}$ from $\vstar$ and keeping the rest in the same order.
\end{update}

For the upper bound instead, given a tree we can find the optimal matching in linear time. We then use this matching for our upper bound $\gbound{\vr{n}}$. For the computation we use the symmetric matrix G which records the maximal relative score of any pair of nodes
\begin{align} \label{eq:Geqn}
G_{ij} = & \smax{\vn_i,\vn_j}{V \setminus \{\vn_i,\vn_j\}}  - \osc{\lo{\vn_i}} - \osc{\lo{\vn_j}}
\end{align}
If we define $\M{\vr{n}}$ to be the optimal matching obtained from $\T{\vr{n}}$, the upper bound is
\begin{align} \label{eq:gbound}
\gbound{\vr{n}} = \sum_{e \in \M{\vr{n}}}G_{e} + \sum_{u\in U}\osc{\lo{u}} 
\end{align}

If the best tree score matches the upper bound $\gbound{\vr{n}}$, the tree network on the dormant nodes is optimal and we do not need to continue expanding the suborder:
\begin{pruning}[Optimality]\label{prune:optimal}
If $\fbound{\vr{n}} = \gbound{\vr{n}}$, we prune $\vr{n}$.
\end{pruning}
A neat corollary is that, if the best network is a matching, we can prove optimality in quadratic time (see Section~\ref{sec:matching}). Finally, if the actual score of the dormant nodes within any of the update rules is larger than $\fbound{\vr{n}}$ but matches the upper bound $\gbound{\vr{n}}$, we can also prune.

\subsection{Computational complexity} \label{sec:complex}

For the complexity considerations, each score computation for a single node can be up to $\bigO(p)$ in our setting. To aid the computations, for each suborder $\vr{n}$ we keep track of the highest score of the remaining dormant nodes $\ghnode \in V \setminus \{\vn_j\}_{j=1}^n$ in the front, further along the order, or completely at the front with a gap: $\osc{\ro{\ghnode,\vn_{n:1}}}$, $\max_{k=1 \ldots n-1} \osc{\ro{\vn_{n:k},\ghnode,\vn_{k-1:1}}}$ and $\osc{\lro{\ghnode}{\vn_{n:1}}}$. Each suborder at stage $n$ is then stored as the ordered list of the $n$ nodes already included and the table of $3(p-n)$ scores.

After stage $(n-1)$, say we have $N_{n-1}$ suborders $O_{n-1}^{N_{n-1}}$ of length $(n-1)$. For each of these we try to append the remaining $(p-n+1)$ nodes to the front for Pruning~\ref{prune:opt_front} and \ref{prune:ord_front}. With the tabulated scores, the optimal front condition can be checked in $\bigO(1)$. Since $N_{n-1} \leq \binom{p}{n-1}$ the total complexity of this pruning is bound by $\bigO(p2^p)$.

For removing right gaps with Pruning \ref{prune:norightgaps}, we can check the condition in $\bigO(p)$ from the tabulated scores and already prune some of the $(p-n+1)$ nodes which could be added to the front. The total complexity is likewise bound by $\bigO(p2^p)$.

The more complicated task is the removing dormant right gaps with Pruning \ref{prune:dormantrightgap} and \ref{prune:dormantrightord}. Here we need the score for each potential dormant node of which there are $\bigO(p)$ and each new score computation is also $\bigO(p)$ in our setting. With the $\bigO(p)$ nodes we try to append to the front this could na\"ively mean a complexity of $\bigO(p^3 N_{n-1})$, however we can avoid this overhead and reduce the complexity by reorganising our algorithm and first running the global pruning, as follows.

For the global Pruning \ref{prune:no_dup}, after the optimal front and right gap pruning we have $N \leq (p-n+1)N_{n-1}$ potential suborders. Of these we remove any duplicates with the same node set, keeping the highest scoring or in numerical order. To compare the sets is $\bigO(pN)$ since we also need to check the suborder set elements. By summing over $n$ the total complexity of this step is bound by $\bigO(p^2 2^p)$. Note that for the equivalent pruning in Section \ref{sec:old_complexity} the complexity was only $\bigO(p2^p)$ since we already know the suborder set elements from the Hasse diagram.

After the global pruning we have $N' \leq \binom{p}{n}$ suborders remaining. For each of these we now update their insertion score tables. For each suborder $\ro{\vn_{n:1}}$ of length $n$, for each of the dormant nodes $\ghnode \in V \setminus \{\vn_j\}_{j=1}^n$ we compute $\osc{\ro{\ghnode,\vn_{n:1}}}$, $\osc{\lro{\ghnode}{\vn_{n:1}}}$ and $\osc{\ro{\vn_{n},\ghnode,\vn_{n-1:1}}}$. Since these are all updates of one node compared to the scores previously computed for $\ro{\vn_{n-1:1}}$, each one takes $\bigO(p)$ and the whole process takes $\bigO(p^2)$. For the maximum score
\begin{align}
& \max_{k=1 \ldots n-1} \osc{\ro{\vn_{n:k},\ghnode,\vn_{k-1:1}}} = \s{\vn_{n}}{\ghnode, \vn_{n-1:1}} + \max_{k=1 \ldots n-1} \osc{\ro{\vn_{n-1:k},\ghnode,\vn_{k-1:1}}}
\end{align}
the maximum on the second line is computed by taking the largest score with the dormant node at the front or further along the order previously computed for  $\vr{n-1}$ and hence is $\bigO(1)$. Updating the score tables therefore is $\bigO(p^2 N')$ and the total complexity is again bound by $\bigO(p^2 2^p)$.

With the updated score tables for the suborders of length $n$, checking the condition for the removal of dormant right gaps with Pruning \ref{prune:dormantrightgap} and \ref{prune:dormantrightord} takes $\bigO(1)$ per dormant node and hence $\bigO(p)$ for each suborder. The total complexity is then bound by $\bigO(p2^p)$. In particular we can check the gap condition as we are updating the score tables.  

For pruning left suborders, the total complexity of Pruning~\ref{prune:opt_back} and \ref{prune:ord_back} is, analogously for the right suborders, bound by $\bigO(p2^p)$. For removing left gaps with Pruning \ref{prune:noleftgaps} we can check the condition in $\bigO(1)$ for each potential new addition at the back by looking up the relevant insertion scores. To check the numerical ordering condition we should additionally track the lowest number of the nodes outside the suborder to avoid any extra computational costs. As for the optimal back pruning, the total complexity is bound by $\bigO(p2^p)$.

For the path-search Pruning \ref{prune:Astar}, computing the lower bound with Prim's algorithm \citep{prim1957shortest} for undirected maximum spanning trees, and also extracting a matching for the upper bound for Pruning \ref{prune:optimal}, is $\bigO(p^2)$. For each update rule for the best suborder, we need to score specific left suborders of the dormant nodes, again with complexity $\bigO(p^2)$. The total complexity of the path-search based pruning and updates is therefore bound by $\bigO(p^2 2^p)$.

The overall worst case total complexity of $\bigO(p^2 2^p)$ is the same as the baseline complexity of Section \ref{sec:old_complexity}. When including the divide-and-conquer rules, we run the above scheme on each potentially disconnected component, but after discovering the optimal network on each component we may need to merge components and rerun. The number of components is limited by $p$, and hence so is the number of times we could merge and rerun. However, the sizes of the components are also smaller than $p$, so for example the scheme where we add one node at a time and rerun would have complexity terms involving $\sum_{k=1}^{p}k^22^k$, again leading to an overall complexity of $\bigO(p^2 2^p)$.  

Including all the pruning rules does not therefore worsen the complexity class of the algorithm. Because we prune suborders, however, we may be able to achieve lower complexity in practice than with the dynamic programming approach. 

\subsection{Polynomial discovery of matchings} \label{sec:matching}

For the lower bound from the maximum spanning tree in Equation (\ref{eq:fbound}) there may be edges with $F_{ij} = 0$. In this case, the best network would not have an edge between $\vn_i$ and $\vn_j$. If the best tree is a matching, we can only have edges between disconnected pairs of nodes. If $\vn_i$ and $\vn_j$ are paired in the matching, then we must have $F_{ik} = 0$ for all $k\neq j$ and $F_{kj} = 0$ for all $k\neq i$. If not the MST would have placed an edge between $\vn_i$ or $\vn_j$ and another node. Note that this would not have created a cycle since the other node would be part of a distinct disconnected pair.

\subsubsection*{Bounds for matchings}

If the best network is a matching, for each connected pair $\vn_i$ and $\vn_j$ we must have 
\begin{align} \label{eq:matching_move}
\osc{\lo{\vn_i, \vn_j}} = \osc{\ro{\vn_i, \vn_j}} 
\end{align}
and
\begin{align} 
\osc{\lo{\vn_j, \vn_i}} = \osc{\ro{\vn_j, \vn_i}} 
\end{align}
since otherwise we could include other parents for one of the nodes, increase the score and no longer have a matching. Combined this is
\begin{align} 
\smax{\vn_i, \vn_j}{V \setminus \{\vn_i,\vn_j\}} & = \osc{\ro{\vn_i, \vn_j}} = \osc{\ro{\vn_j, \vn_i}}
\end{align}
Looking at one of the nodes
\begin{align} \label{eq:smatchnode}
\osc{\lo{\vn_i}} = & {} \osc{\lo{\vn_i, \vn_j}} - \osc{\lo{[\vn_i], \vn_j}} = \osc{\ro{\vn_i, \vn_j}} - \osc{\ro{\vn_j}} 
\end{align}
comparing to Equations (\ref{eq:Feqn}) and (\ref{eq:Geqn}) we have
\begin{align} 
G_{ij} & = -F_{ij}
\end{align}

On the other hand, if $\vn_i$ and $\vn_k$ are disconnected in the matching, 
\begin{align} 
\smax{\vn_i, \vn_k}{V \setminus \{\vn_i,\vn_k\}} & = \osc{\lo{\vn_i}} + \osc{\lo{\vn_k}}
\end{align}
since their relative order does not matter. Therefore
\begin{align} 
G_{ik} & = 0 = -F_{ik}
\end{align}
and overall $G = -F$ when the best network is a matching.

\subsubsection*{Equality for matchings}
Since $G$ and $F$ are opposite of each other, the minimum spanning tree for $G$ is a maximum spanning tree for $F$, and so the same matching is an MST for $G$. As a final step, we need to show equality of the lower and upper bounds:
\begin{align} \label{eq:boundmatch}
& \fbound{\vr{n}} = \sum_{e \in \M{\vr{n}}}F_{e} + \sum_{\vn\in V}\osc{\ro{\vn}} \nonumber \\
= & \gbound{\vr{n}} = \sum_{e \in \M{\vr{n}}}G_{e} + \sum_{\vn\in V}\osc{\lo{\vn}}
\end{align}

\begin{proof}
For any disconnected node in the matching, say $\vn_{k}$, including any additional parents cannot improve the score, since one could include edges to those parents and have a tree, contradicting that our best tree is a matching. For disconnected nodes we have
\begin{align} 
\osc{\ro{\vn_k}} = \osc{\lo{\vn_k}}
\end{align}
and no contributions from $F$ or $G$. Otherwise, for connected pairs, $\vn_i$ and $\vn_j$, we have
\begin{align} 
& F_{ij} + \osc{\ro{\vn_i}} + \osc{\ro{\vn_j}} \nonumber \\
= {} & \osc{\ro{\vn_i,\vn_j}} = \osc{\lo{\vn_i,\vn_j}} \nonumber \\
= {} & G_{ij} + \osc{\lo{\vn_i}} + \osc{\lo{\vn_j}}
\end{align}
from the definition in Equations (\ref{eq:Feqn}) and (\ref{eq:Geqn}), and the result in Equation (\ref{eq:matching_move}). The terms agree for disconnected nodes and pairs of connected nodes in the matching, and so agree for the entire matching.  \end{proof}

\subsubsection*{Polynomial optimality testing}
In computing the score tables for $F$ to find the best tree, we only need to consider each node with no, or one parent, and building the matrix $F$ can be done in quadratic time for usual scores like the BDe or BGe \citep{heckerman1995learning,geiger2002parameter,kuipers2014addendum}. Finding the best tree network is well-known to be polynomial \citep{chow1968approximating}, and we can do so from $F$ in quadratic time \citep{prim1957shortest}.

For the matrix $G$, however, we need the best scoring parent set for each node, and the best subset selection problem is generally NP-hard \citep{natarajan1995sparse}. With a restricted search space \citep{kuipers2022efficient}, or a restricted in-degree \citep{friedman2003being}, computing the score tables for $G$ becomes polynomial.

Finding the optimal DAG is NP-hard even for a restricted in-degree (of at least 2, \citealp{chickering1996learning}, or at least 3, \citealp{chickering2004large}), so we consider that case where constructing $G$ is still polynomial. Finding the upper bound from $G$ is linear and can likewise be computed in polynomial time.

Overall, if the best causal Bayesian network in a search space with restricted in-degree is a matching then things simplify. We can find it (from $F$) and prove its optimality (from $G$) in polynomial time, and, once the score tables have been computed, in quadratic time. 

\clearpage

\subsection{Supplementary Figures}

\begin{figure*}[h]
\centering
\begin{tabular}{c}
\includegraphics[width=\textwidth]{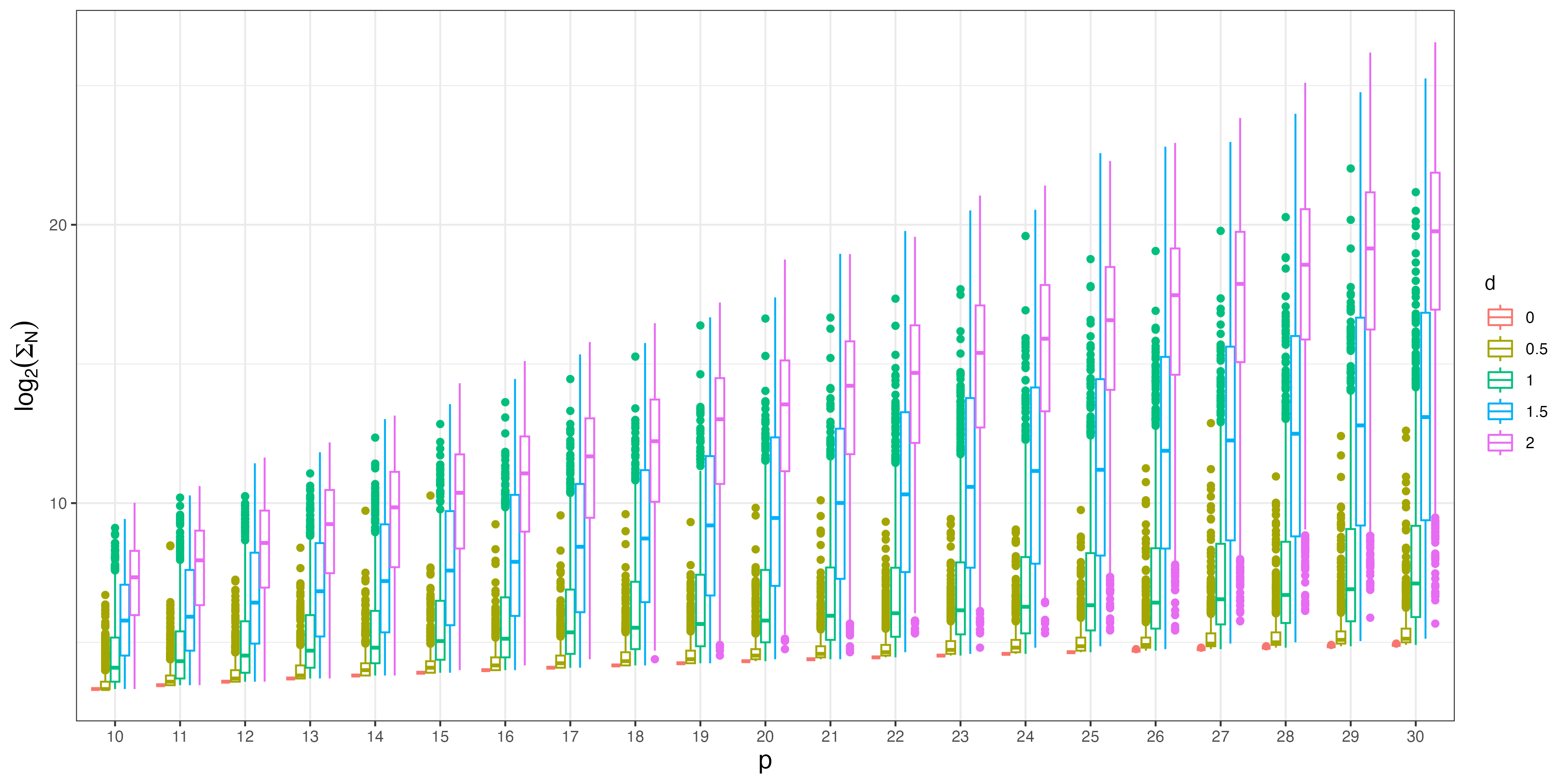}\\
\includegraphics[width=\textwidth]{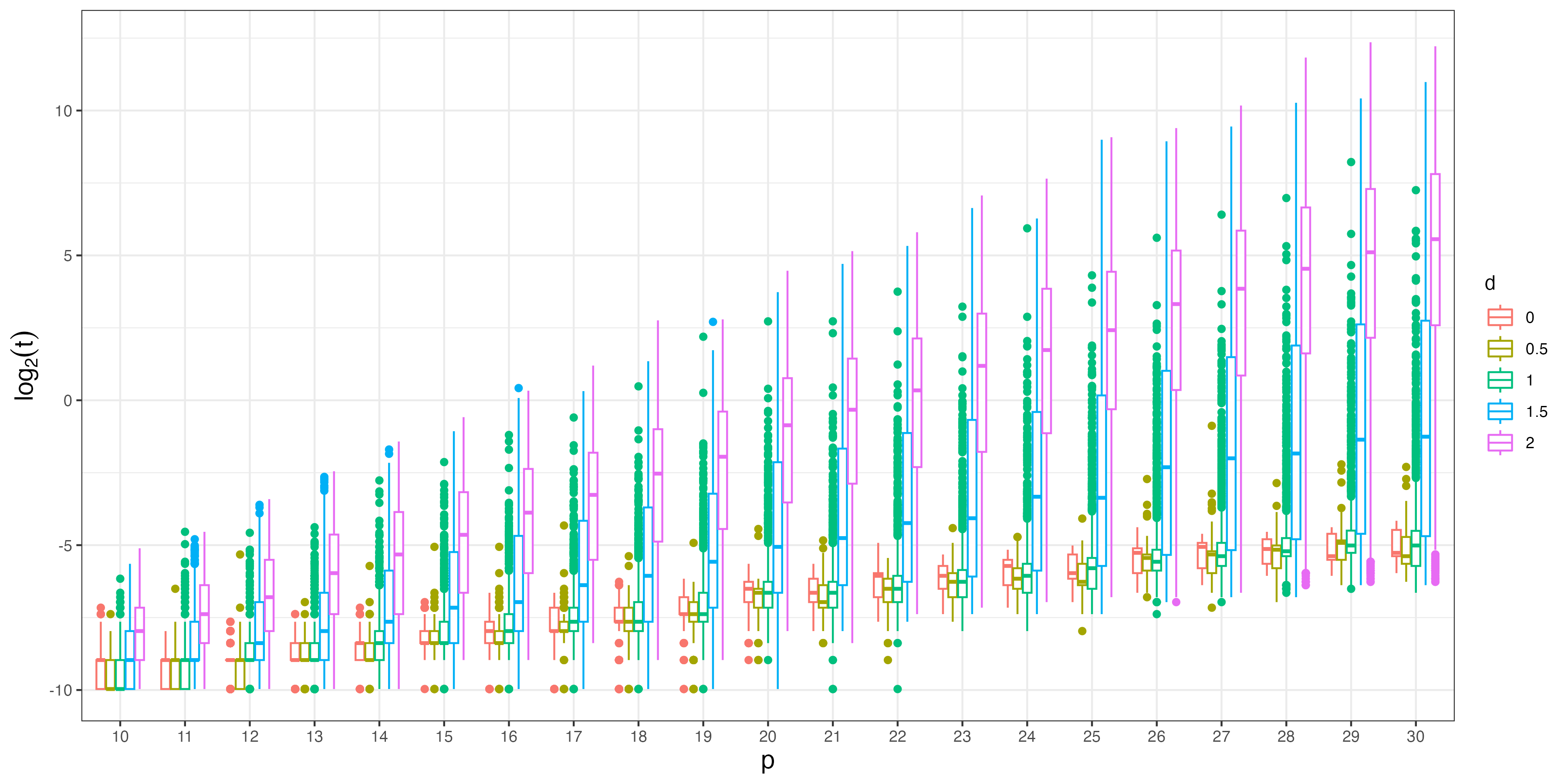}
\end{tabular}
\caption{More complete version of \fref{fig:boxplots} showing the logarithm of the total number of suborders $\Sigma_N$ and  runtime $t$ (in seconds) for all the network sizes $p \in \{10,\ldots,30\}$, stratified by the network density $d$.}
\label{fig:boxplots_wide}
\end{figure*}

\begin{figure*}[t]
\includegraphics[width=\textwidth]{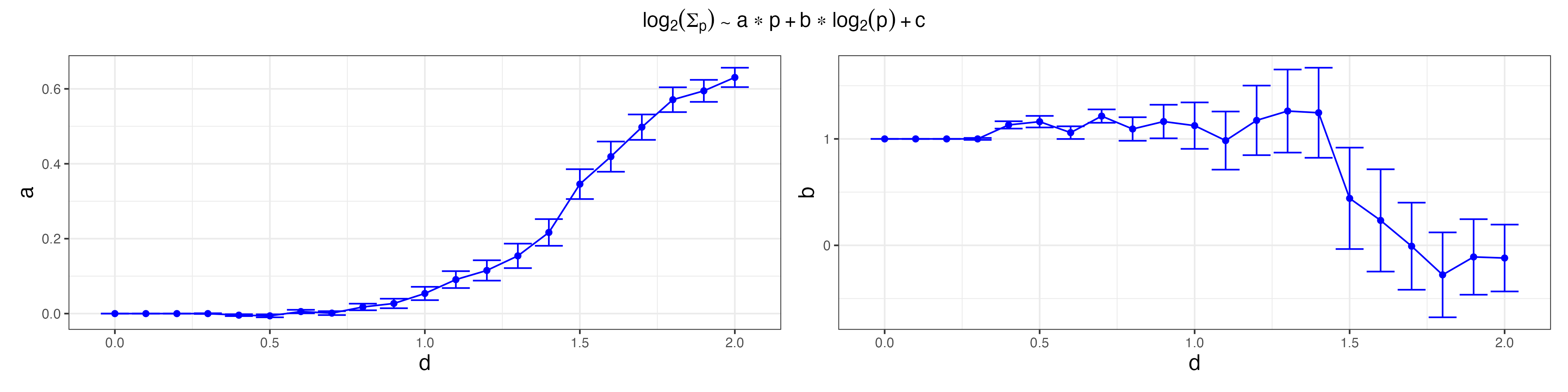}
\caption{The regression coefficients for a median regression of the logarithm of the total number of suborders $\log_{2}(\Sigma_N)$ against $p$ and $\log_2(p)$, with standard errors, as a function of the network density $d$.}
\label{fig:regrplots_med}
\end{figure*}

\begin{figure*}[bht]
\centering
\begin{tabular}{cc}
\centering
\includegraphics[width=0.49\textwidth]{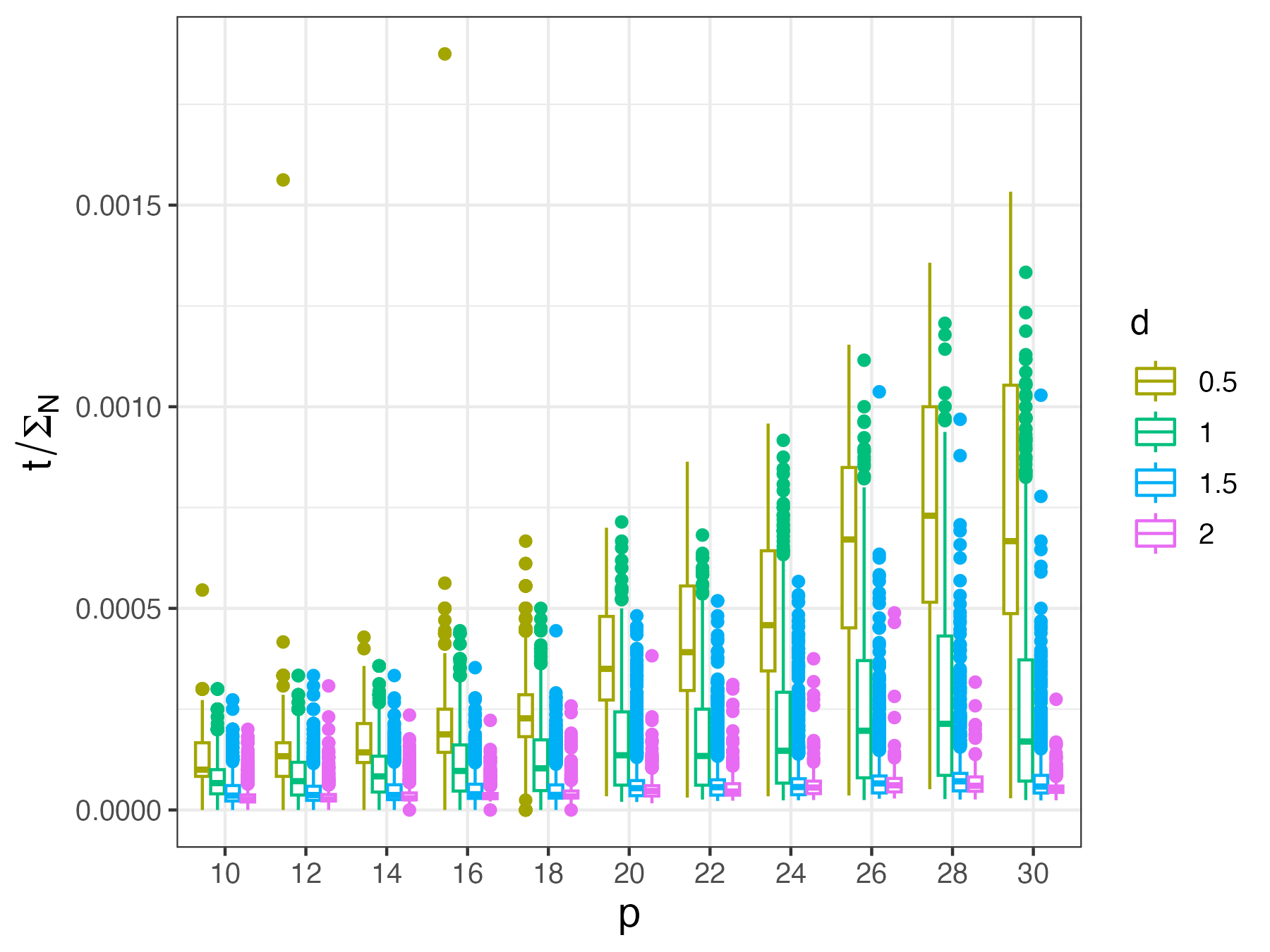}
&
\includegraphics[width=0.49\textwidth]{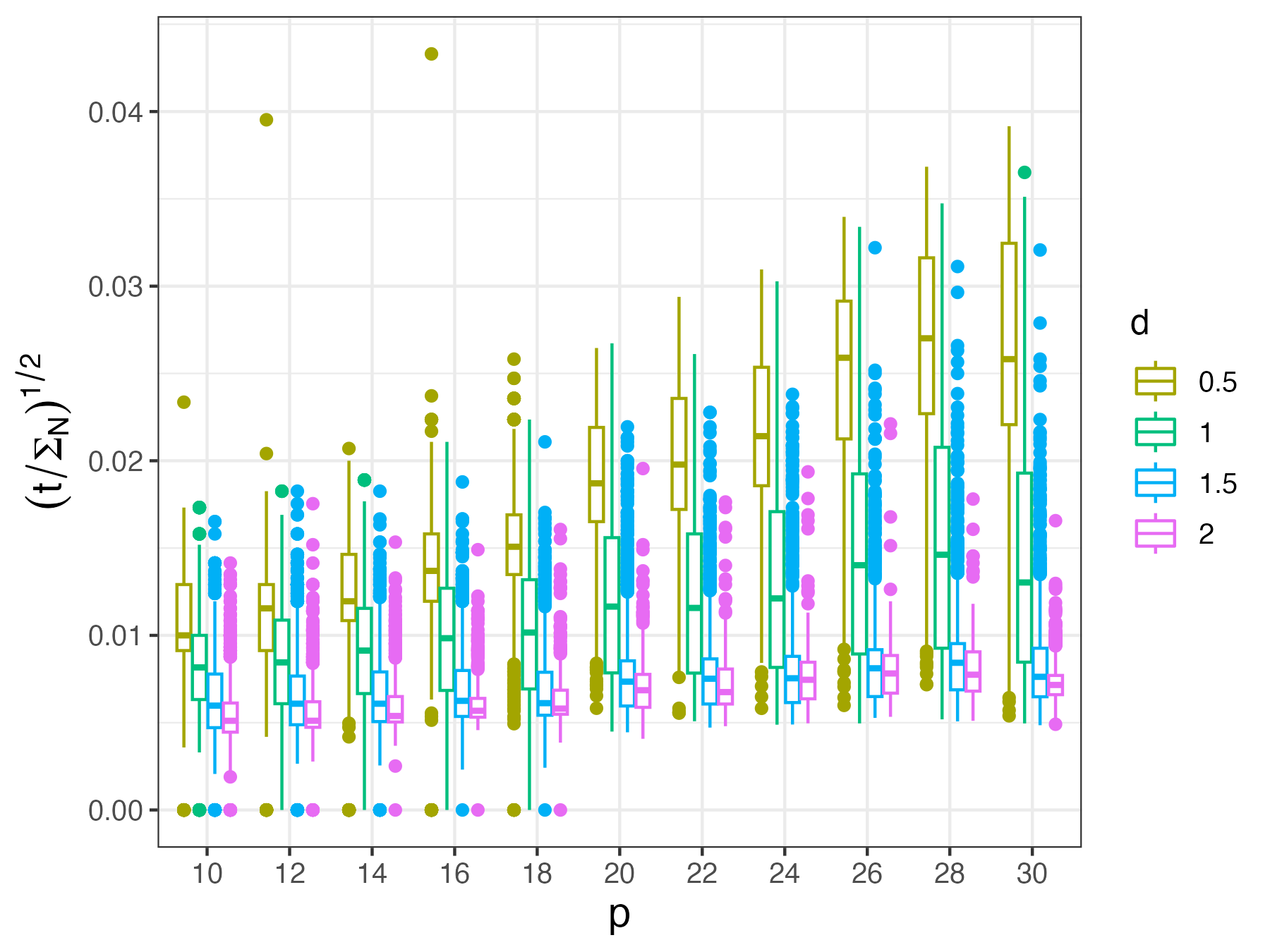}
\end{tabular}
\caption{The ratio of the total runtime to the total number of suborders: $t/\Sigma_N$ for networks of different density $d$. We would expect up to quadratic behaviour in $p$, though it appears more linear for higher densities. With the square root transform (right panel), the apparent sub-linear behaviour for higher densities suggests an overall sub-quadratic dependency.}
\label{fig:ratio_boxplot}
\end{figure*}

\clearpage

\begin{figure*}[t]
\centering
\includegraphics[width=0.6\textwidth]{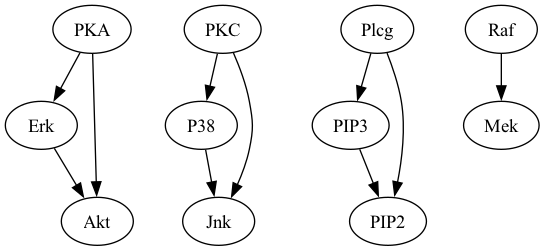}
\caption{The optimum network learned from the Sachs dataset \citep{sachs2005causal}, which aligns with that learned with most causal discovery algorithms \citep{rios2021benchpress}. Each component is fully connected and hence score-equivalent to any acyclic redirecting of the edges.}
\label{fig:Sachs_net}
\end{figure*}

\begin{figure*}[bh]
\vspace{4cm}
\centering
\begin{tabular}{cc}
\includegraphics[width=0.49\textwidth]{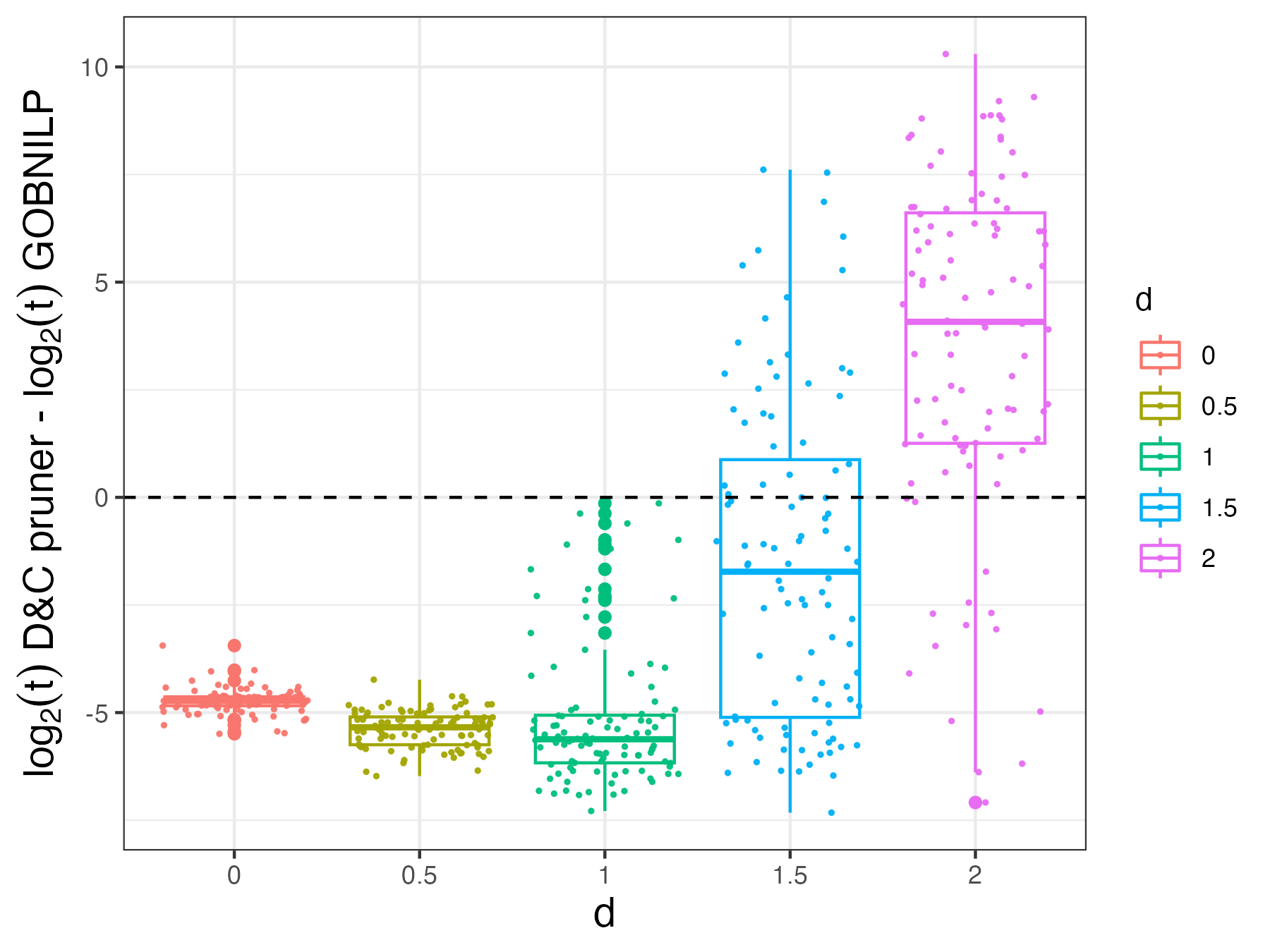}
&
\includegraphics[width=0.49\textwidth]{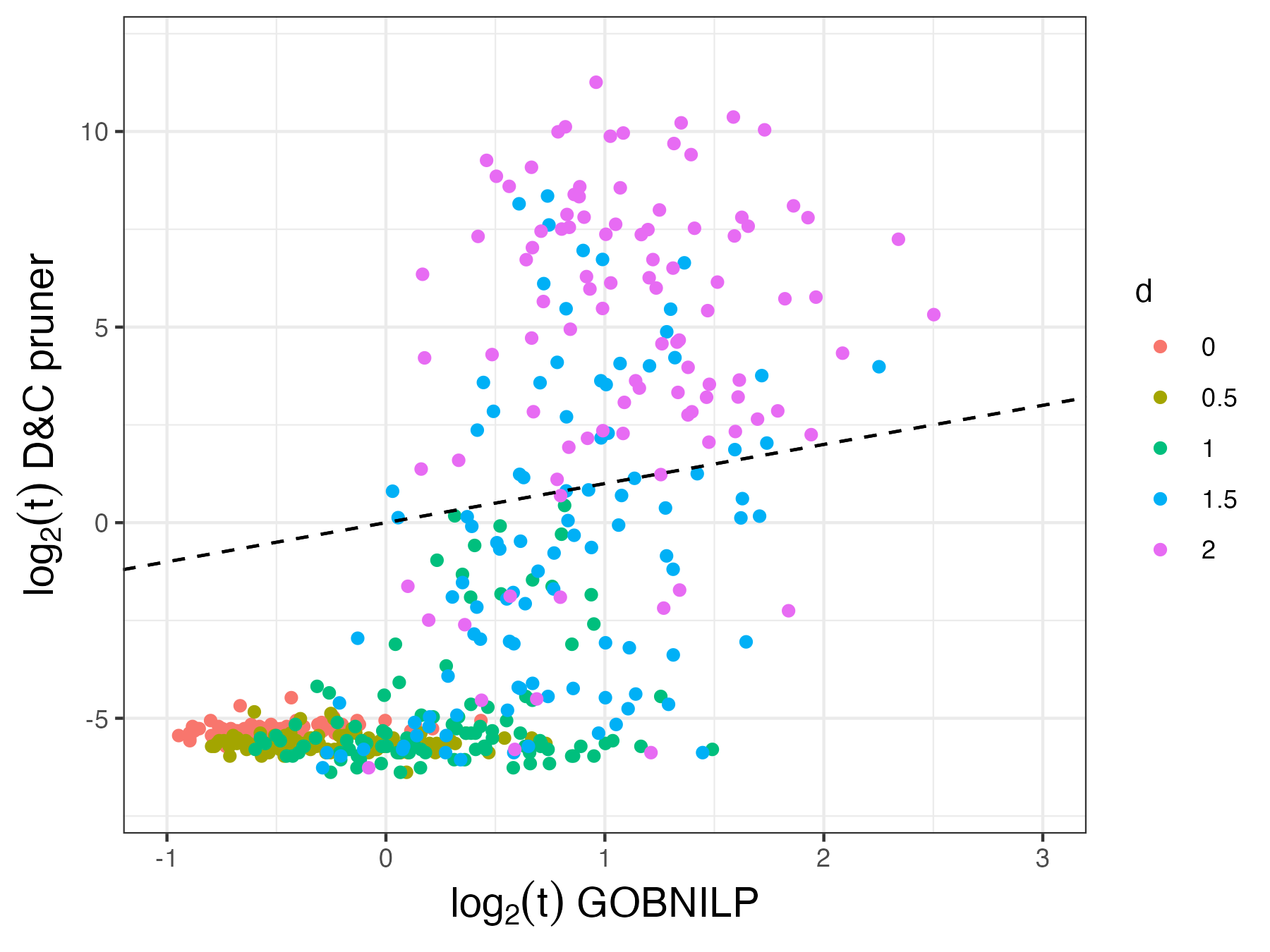}
\end{tabular}
\caption{The logarithm of the relative runtime of our approach (D\&C pruner) compared to GOBNILP for networks of size $p=30$ along with the comparison of the log runtimes (in seconds) of the two approaches, coloured by density.}
\label{fig:vsGOBNILP}
\end{figure*}

\clearpage

\subsection{Supplementary Algorithms}\label{sec:algos}

\begin{algorithm}[h]
\DontPrintSemicolon

\hrulefill
\caption{mainTransition}\label{alg:trans}
\vspace{-1.5ex}
\hrulefill

\KwData {$M_{n-1}$ list of orders of length $n-1$ }
%\Ensure $y = x^n$

$M_{n} \gets \{\}$ \;

\For{$ \bar m = \vr{n-1} \in M_{n-1}$ \tcc*[r]{$\bigO({p \choose {n-1}})$}}{
    $R \gets$ $noRightGaps(\bar m)$ \tcc*[r]{$\bigO(p)$} \;
    \For{$ j  \in V \setminus \{\vn_{j}\}_{j=1}^{n-1}$ \tcc*[r]{$\bigO(p)$}} { 
        
        \If{$j \notin R$ \tcc*[r]{$\bigO(1)$}} { 

            \textbf{continue} \;  % \Comment{eq. to pruning?}
        }
        \If{!$optimalFront(\bar m, j)$ \tcc*[r]{$\bigO(1)$}}{ %
            \textbf{continue} \; % \Comment{eq. to pruning?}
        }
        \If{$!orderedFront(\bar m, j)$}{
            \textbf{continue} \;
        }
        
        $\bar u \gets \ro{j,\vn_{n-1:1}}$ \;
        
        Compute $\osc{\bar u}$ \; \tcc*[r]{$\bigO(1)$} 
        $M_{n} \gets M_{n} \cup \{\bar u\}$  \;
    }
}

$M_{n} \gets$ $pruneEqualSets(M_{n}$) \tcc*[r]{$\bigO({p \choose n}n \times p)$}\;
\For{$ \bar m \in M_n$ \tcc*[r]{$O({p \choose n})$}}{ 

    Update insertion scores for $\bar m$\; \tcc*[r]{$\bigO(p^2)$}
}
\For{$ \bar m \in M_n$ \tcc*[r]{$\bigO({p \choose n})$}}{ 
    \If{$hasDormantGap(\bar m$) \tcc*[r]{$\bigO(p)$} } { 
        $M_{n} \gets M_{n} \setminus \{\bar m\}$ \;
        }

    \If{$!orderedDormantGap(\bar m$) \tcc*[r]{$\bigO(p)$}}{ 
        $M_{n} \gets M_{n} \setminus \{\bar m\}$ \;
    }
}

\Return $M_n$ \;

\vspace{-1.5ex}
\hrulefill
\end{algorithm}

\begin{algorithm}
\DontPrintSemicolon

\hrulefill
\caption{noRightGaps (\textit{c.f.}~Pruning~\ref{prune:norightgaps})}
\vspace{-1.5ex}
\hrulefill

\KwData{ A right order $\bar m = \vr{n-1}$  }
$R = \{\}$ \;

$m \gets -1$ \;

\For{$\ghnode \in V \setminus \{\vn_{j}\}_{j=1}^{n-1} $}{
    \If{$\osc{\lro{h}{\vn_{n-1:1}}} = \osc{\ro{h,\vn_{n-1:1}}}$}{
        \If{$h > m$}{
            $m \gets h$ \;
        }
    }
}

\For{$\ghnode \in V \setminus \{\vn_{j}\}_{j=1}^{n-1} $}{
    \If{$\ghnode \ge m$}{
        $R \gets R \cup \{\ghnode \}$ \;
    }
}
\Return $R$

\vspace{-1.5ex}
\hrulefill
\end{algorithm}

\begin{algorithm}
\DontPrintSemicolon

\hrulefill
\caption{optimalFront (\textit{c.f.}~Pruning~\ref{prune:opt_front})}
\vspace{-1.5ex}
\hrulefill

\KwData{A right order $\bar m = \vr{n-1}$}
\KwData{A new front node $j$ }
\KwData{Pre-calculated max insert scores for \(\bar m\)}
%\Ensure $y = x^n$
$m \gets \osc{\ro{j,\vn_{n:1}}}$ \;

\(s \gets \max_{k\le n} \osc{\ro{\vn_{n:k},j,\vn_{k-1:1}}}\) \; \tcc*[r]{$\bigO(1)$}

    \If{\(s > m\)}{
        \Return False \;
    }
\Return True

\vspace{-1.5ex}
\hrulefill
\end{algorithm}

\begin{algorithm}
\DontPrintSemicolon

\hrulefill
\caption{orderedFront (\textit{c.f.}~Pruning~\ref{prune:ord_front})}
\vspace{-1.5ex}
\hrulefill

\KwData{A right order $\vr{n}$  }
%\Ensure $y = x^n$
 \If{\(\osc{\vr{n}} = \osc{\ro{\vn_{n-1},\vn_n,\vn_{n-2:1}}}\)}{
    \If{\(\vn_{n} < \vn_{n-1}\)}{
        \Return True\;
}
}
\Return False

\vspace{-1.5ex}
\hrulefill
\end{algorithm}

\begin{algorithm}
\DontPrintSemicolon

\hrulefill
\caption{hasDormantGap (\textit{c.f.}~Pruning~\ref{prune:dormantrightgap})}
\vspace{-1.5ex}
\hrulefill

\KwData{A right order $\bar m = \vr{n}$}
%\Ensure $y = x^n$
\KwData{Pre-calculated max insert scores for \(\bar m\).}
    $m \gets \osc{\lro{\ghnode}{\vn_{n:1}}}$ \; \tcc*[r]{$\bigO(1)$}

    \For{$ \ghnode \in V \setminus \{\vn_{j}\}_{j=1}^{n}$ \tcc*[r]{$\bigO(p)$}} { 

    \(s \gets  \max_{k} \osc{\ro{\vn_{n:k},\ghnode,\vn_{k-1:1}}}\)\; \tcc*[r]{$\bigO(1)$}
    
        \If{\(s > m\)}  {
            \Return True
        }
    }
%\EndFor
\Return False

\vspace{-1.5ex}
\hrulefill
\end{algorithm}

$\left.\right.$
\vspace{1cm}

\begin{algorithm}[h]
\DontPrintSemicolon

\hrulefill
\caption{orderedDormantGap (\textit{c.f.}~Pruning~\ref{prune:dormantrightord})}
\vspace{-1.5ex}
\hrulefill

\KwData{A right order $\bar m = \vr{n}$}

\KwData{Pre-calculated max insert scores for \(\bar m\)}

    $m \gets \osc{\lro{\ghnode}{\vn_{n:1}}}$ \; \tcc*[r]{$\bigO(1)$}

    \For{$ \ghnode \in V \setminus \{\vn_{j}\}_{j=1}^{n}$ \tcc*[r]{$\bigO(p)$}}{ 

    \(s \gets  \osc{\ro{\vn_{n},\ghnode,\vn_{n-1:1}}}\) \; \tcc*[r]{$\bigO(1)$} 
   
        \If{\(s = m\) $\And$ \(\ghnode < \vn_n\)}  {
            \Return True
        }

}
\Return False

\vspace{-1.5ex}
\hrulefill
\end{algorithm}

\begin{algorithm}[h]

\DontPrintSemicolon

\hrulefill
\caption{pruneEqualSets (\textit{c.f.}~Pruning~\ref{prune:no_dup})}\label{alg:equal_prune}
\vspace{-1.5ex}
\hrulefill

\KwData{\(A \in \{0,1\}^{N\times p}\) an \(N\times p\) binary row matrix indicating elements in orders}
\KwData{\(order\_scores \in \textbf R^N\)}
\(start\_inds \gets [1,1,2,\dots,N]\) \;  \tcc*[r]{First element (1) indicates the column to be used in \(A\)}

\(L=\{\}\) \; \tcc*[r]{List of indices for the maximal scoring order of same sets}

\(Q=\{\}\)\;  \tcc*[r]{Indices to consider}

\(Q.push\_back(start\_inds)\)\;

\While{\(Q \ne \{\}\)}{
    \(inds \gets Q.pop\_back()\) 

    \(k \gets inds[1]\) \; 
    
    \(one\_inds \gets [k+1]\)\;  \tcc*[r]{To be used in col \(k+1\) of \(A\)}
    \(zero\_inds \gets [k+1]\)\;
    
    \For{\(i \in inds\)}{
        \If{\(A_{ik} = 0\)}{
                \(zero\_inds.push\_back(i)\)\;
            }{

                \(one\_inds.push\_back(i)\)\;
            }
    }

    \eIf{\(k=p\) \tcc*[r]{Last column, where scores are used}}{ 
        \If{\(zero\_inds.size() \ne 0\)}{
            \(i^* \gets \underset{i \in zero\_inds}{\mathrm{\arg\max}}\,order\_scores[i]\) \;
            
            \(L.push\_back(i^*)\) \;
}
        
        \If{\(one\_inds.size() \ne 0\)}{
            \(i^* \gets \underset{i \in one\_inds}{\mathrm{\arg\max}}\,order\_scores[i]\) \;
            
            \(L.push\_back(i^*)\) \;
        }
        }{
 
         \If{\(zero\_inds.size() > 1\)}{
            \(Q.push\_back(zero\_inds)\) \;
}
         \If{\(one\_inds.size() > 1\)}{
             \(Q.push\_back(one\_inds)\)\;
}
}
}
\Return $L$

\vspace{-1.5ex}
\hrulefill
\end{algorithm}

\end{document}